\documentclass[a4paper,preprint,nofootinbib]{revtex4-1}

\usepackage[utf8]{inputenc}
\usepackage{graphicx}
\usepackage{mathtools}
\usepackage{amsfonts}
\usepackage{chemmacros}
\usepackage{csquotes}
\usepackage{mhchem}
\usepackage[labelformat=empty,caption=false]{subfig}
\usepackage{siunitx}

\usepackage{hyperref}

\usepackage[
    capitalize,
]{cleveref}

\begin{document}

\title{Quantification of electron correlation effects -- Quantum Information Theory versus Method of Increments}

\author{Christian Stemmle}
\email{christian.stemmle@fu-berlin.de}
\author{Beate Paulus}
\affiliation{Institut f\"ur Chemie und Biochemie - Takustr. 3, 14195 Berlin, Freie Universit\"at Berlin, Germany}

\begin{abstract}
Understanding electron correlation is crucial for developing new concepts in electronic structure theory, especially for strongly correlated electrons. We compare and apply two different approaches to quantify correlation contributions of orbitals: Quantum Information Theory (QIT) based on a Density Matrix Renormalization Group (DMRG) calculation and the Method of Increments (MoI).
Although both approaches define very different correlation measures, we show that they exhibit very similar patterns when being applied to a polyacetelene model system.
These results suggest one may deduce from one to the other, allowing the MoI to leverage from QIT results by screening correlation contributions with a cheap (``sloppy'') DMRG with a reduced number of block states.
Or the other way around, one may select the active space in DMRG from cheap one-body MoI calculations.
\end{abstract}

\maketitle

\section{Introduction}
The objective of electronic structure theory is finding sufficient accurate solutions to the Schrödinger equation for chemical or physical applications. Unfortunately the correlated motion of this many-body problem turns out hard to be solved. An approach yielding numerically exact solutions is long known with the Full Configuration Interaction (FCI) method \cite{Lowdin-1957}, usually based on a Hartree-Fock (HF) \cite{Hartree-1928,Fock-1930,Slater-1930} calculation which is missing electron correlation effects. However FCI is unfeasible for large systems due to factorial scaling with the number of electrons involved. The main challenge of electronic structure theory thus remains as finding approximate solutions, yielding results within a defined error margin.\\

Various approaches provide different trade-offs between accuracy and computational scaling. Density Functional Theory (DFT) \cite{Hohenberg-1964} established itself for being capable of dealing with most systems of interest at reasonable computational cost, but has problems describing systems involving strong correlations and lacks systematic improvability.
Additionally DFT requires some a priori knowledge of data, to validate whether a certain functional is suitable for the system in question. In other words, although DFT calculations may achieve high precision (i.e. have small statistical error), they may lack accuracy (i.e. have large systematic error), making them questionable for predictions.\\

Wave function based methods on the other hand are systematic truncations to the FCI problem, allowing them to treat any system with any desired accuracy and precision at the cost of higher computational effort. It is thus highly desirable to find superior trade-offs, making large system accessible for wave function based methods.
Coupled-Cluster theory \cite{Bartlett-1981} evolved to be the standard method for \emph{single-reference} systems, i.e. systems where the major part of the electron-electron interactions are accounted for by HF and thus most electron correlation effects are sufficiently describe by including \emph{dynamical} correlations, by creating excitations based on the HF solution.
Systems inadequately described by such \emph{single-reference} approaches are called \emph{strongly correlated} and require \emph{multi-reference} treatments.
A common approach for describing strong correlations are multi-configurational methods, like Multi Configuration Self Consistent Field (MCSCF) and its variant Complete Active Space Self Consistent Field (CASSCF) \cite{Roos-1980,Szalay-2012}, including selected configurations of same or similar contributions to the wave function, while the HF method only considers a single configuration.
Dynamic correlation can then be added in a second step by generating excitations based on many reference configurations, hence the name \emph{multi-reference} methods.\\

The previously mentioned active space in CASSCF is a selection of a small subset of orbitals, which are deemed most relevant for static correlation. Hence this method still suffers from the same unfortunate scaling as the FCI problem, only less severe since usually the orbital space is much smaller. For cases where larger active spaces are required the Density Matrix Renormalization Group (DMRG) \cite{White-1992} is a suitable  alternative. In DMRG the diagonalization of a large Hamiltonian is sidestepped by chunking it into many smaller matrices and keeping only the most important contributions. Another approach primarily developed for periodic and extended systems is the Method of Increments (MoI) \cite{Stoll-1992,Stoll-1992a}, where the correlation energy is expanded in terms of groups of occupied orbitals (called centers). Both, DMRG and MoI, can benefit from localized orbitals to reduce long range correlations, which leads to faster convergence and reduced computational cost.\\

Another interesting aspect common to DMRG and MoI is the possibility to analyze and quantify the origin of correlation effects. While in the MoI the obtained increments can be directly interpreted, we can apply Quantum Information Theory (QIT) \cite{Horodecki-2009,Szalay-2015} to extract this data from the DMRG wave function (or in general any correlated wave function). Although there are technical differences between both, as will be explained in \cref{sec:theory}, we will show in \cref{sec:results} that we indeed find very similar trends. The model and computational details are presented in \cref{sec:models} and \cref{sec:compdetails} respectively. The conclusion follows in \cref{sec:summary}.

\section{Theory}
\label{sec:theory}

\subsection{Method of Increments}
In the Method of Increments (MoI) \cite{Stoll-1992,Stoll-1992a,Paulus-2006,Voloshina-2014,Fertitta-2015} one partitions the correlation energy into contributions by assigning the occupied orbitals to different centers. These centers are correlated with the rest of the system (virtual orbitals) at different levels. The first level are the \emph{one-center} or \emph{one-body} increments
\begin{equation}
    \varepsilon_i = E_i - E_{HF}
\end{equation}
where $E_{HF}$ is the reference energy of the uncorrelated system (Hartree-Fock), and $E_i$ the energy obtained by correlating center $i$ only, using a size extensive correlation method chosen by the user. Further levels are calculated by combining two, three or more centers and obtaining their total energies $E_{ij}$, $E_{ijk}$ and so on. Again, only the correlation energies
\begin{align}
    \varepsilon_{ij} &= E_{ij} - E_{HF} \\
    \varepsilon_{ijk} &= E_{ijk} - E_{HF}
\end{align}
are considered. To avoid double accounting for correlations from lower levels, these have to be subtracted from the higher levels to obtain the corresponding increments
\begin{align}
    \Delta \varepsilon_{ij} &=  \varepsilon_{ij} - \varepsilon_i - \varepsilon_j \label{eq:2center} \\
    \Delta \varepsilon_{ijk} &= \varepsilon_{ijk} - \Delta \varepsilon_{ij} - \Delta \varepsilon_{ik} - \Delta \varepsilon_{jk} - \varepsilon_i - \varepsilon_j  - \varepsilon_k.
\end{align}
The correlation energy of the whole system can then be expanded in a series of increments
\begin{equation}
    E_{\rm corr} = \sum_i \varepsilon_i + \sum_i\sum_{j>i}\Delta\varepsilon_{ij} + \sum_i\sum_{j>i}\sum_{k>j}\Delta\varepsilon_{ijk} + \dots
\end{equation}
The number of increments at each level increases combinatorially making the method expensive at higher levels. But as higher level contributions converge to zero the expansion can be truncated, usually after the third level. Furthermore, using localized occupied orbitals allows neglecting contributions for pairs (or groups) of orbitals which are spatially separated, reducing the number of increments to be calculated further.\\

The individual correlation energies $\varepsilon_i$, $\varepsilon_{ij}$, $\varepsilon_{ijk}$, \emph{etc.} may also be interpreted directly as the correlation effects arising from each center, while increments like $\Delta\varepsilon_{ij}$ or $\Delta\varepsilon_{ijk}$ measure the additional effect of correlating the combined group.

\subsection{Density Matrix Renormalization Group (DMRG) and Quantum Information Theory (QIT)}

The Density Matrix Renormalization Group (DMRG) method was first invented by \textcite{White-1992} and is suitable to treat strong correlations in large active spaces and is closely connected to Quantum Information Theory (QIT) \cite{Horodecki-2009} which allows for quantification of correlation effects in terms of (groups of) orbitals. As both, DMRG and QIT, are described in detail in various reviews \cite{Legeza-2008,Marti-2010,Kurashige-2013,Wouters-2014,Szalay-2015,Olivares-Amaya-2015,Chan-2016}, we restrict ourselves to a brief summary here.\\

In the Density Matrix Renormalization Group (DMRG) method the Full Configuration Interaction (FCI) wave function is approximated by trying to find its most important contribution. In essence the complete FCI Hamiltonian is not diagonalized at once, but optimized iteratively in subspaces of e.g. 2 orbitals. During this iterative scheme the most important contributions are kept and carried over to the next iteration. This procedure is facilitated by storing the wave function in a Matrix Product State (MPS), whose accuracy can be controlled by the dimensions connecting two matrices (called \emph{number of block states} or \emph{virtual dimensions}). Both, the iterative diagonalization of smaller subspaces and efficient storing of the wave function in MPS format, allow for treatment of large active spaces and is especially suitable for treating static (or strong) correlations.

For the MPS the orbitals are thought of to be arranged in a linear chain, of arbitrary order. The Configuration Interaction (CI) wave function is then given by
\begin{equation}
    |\Psi\rangle = \sum_{\alpha_1,\dots,\alpha_d} U(\alpha_1,\dots,\alpha_d) |\phi_{\alpha_1}^{\{1\}}\rangle\otimes\dots\otimes|\phi_{\alpha_d}^{\{d\}}\rangle
\end{equation}
where ${\{i\}}$ indicates the orbitals position on the chain of length $d$. The indices $\alpha_i$ label the single-orbital basis states $|\phi_{\alpha_i}^{\{i\}}\rangle$, which correspond to the $q=4$ spin occupations of a spatial orbital: $|\phi^{(1)}_\alpha\rangle \equiv |-\rangle$,  $|\phi^{(2)}_\alpha\rangle \equiv |\downarrow\rangle$,  $|\phi^{(3)}_\alpha\rangle \equiv |\uparrow\rangle$ and $|\phi^{(4)}_\alpha\rangle \equiv |\uparrow\downarrow\rangle$. The CI coefficients are given by the $d$-order tensor $U(\alpha_1,\dots,\alpha_d)$ and is capable of storing coefficients for any configuration with an electron count from 0 to $2d$. Thus its memory requirement grows exponentially as $d^q$.\\

The Matrix Product State (MPS) facilitates these memory requirements by factorizing the tensor $U(\alpha_1,\dots,\alpha_d)$ as a product of low order tensors and with
controlled rank
\begin{equation}
    U(\alpha_1,\dots,\alpha_d) = \boldsymbol{A}_1(\alpha_1)\boldsymbol{A}_2(\alpha_2)\cdots\boldsymbol{A}_{d-1}(\alpha_{d-1})\boldsymbol{A}_d(\alpha_d)
\end{equation}
where each matrix $\boldsymbol{A}_i(\alpha_i)$ corresponds to one molecular orbital $i$. The factorization can formally be obtained by succeeding application of singular value decompositions and is exact, i.e.~the full tensor $U(\alpha_1,\dots,\alpha_d)$ can be recovered without loss of information. As a result the size of the matrices is still growing exponentially towards the center of the chain \cite{Schollwock-2005}.
For the MPS to reduce the memory requirements, one approximates the matrices $\boldsymbol{A}_i(\alpha_i)$ by defining an upper limit to the matrix dimensions called \emph{number of block states} or \emph{virtual bond dimensions} $M$. In practice the challenge of DMRG is now to find an appropriate value of $M$ and a suitable order of orbitals in the chain which will lead to an optimized set of matrices $\boldsymbol{A}_i(\alpha_i)$ to describe the CI wave function.\\

QIT is closely connected to DMRG, as it allows for quantification of orbital correlation contributions (entanglement) based on the CI wave function. In turn the QIT results may be used to find better parameters for the DMRG calculation.
One can easily calculate the $n$-orbital reduced density matrices, by contracting the MPS over all but $n$ orbitals, e.g. the \emph{one-orbital} reduced density matrix is given by
\begin{widetext}
\begin{align}
    \rho_i(\alpha_i,\alpha_i^\prime) &= \mathrm{Tr}_{1,\dots,\not{i},\dots,d}|\Psi\rangle\langle\Psi| \\
    &= \sum_{\alpha_1,\dots,\not{\alpha_i},\dots,\alpha_d} U(\alpha_1,\dots,\alpha_i,\dots,\alpha_d) \overline {U(\alpha_1,\dots,\alpha_i^\prime,\dots,\alpha_d)}.
\end{align}
\end{widetext}

The correlation contribution of a single orbital $i$ are then given by the \emph{one-orbital} von Neumann entropy \cite{Legeza-2003b}
\begin{equation}
    S_i = -\mathrm{Tr}\left(\rho_{i}\ln\rho_{i}\right) = -\sum_\alpha \omega_{i,\alpha} \ln \omega_{i,\alpha} \label{eq:S1},
\end{equation}
where $\omega_{i,\alpha}$ are the eigenvalues of the \emph{one-orbital} reduced density matrix $\rho_i(\alpha,\alpha^\prime)$. The above equation will give small values if all main configurations of the CI wave function have the same occupation in orbital $i$, i.e.~one of the $\omega_{i,\alpha}$ will be close to $1$ while the other three $\omega_{i,\alpha}$ are negligible.
Largest values are obtained if the occupation in orbital $i$ for all main configurations is equally distributed over all $\alpha_i$ ($|-\rangle$, $|\downarrow\rangle$, $|\uparrow\rangle$, $|\uparrow\downarrow\rangle$), i.e.~when $\omega_{i,\alpha}=0.25$ for all $\alpha$, then $S_i=-4\times0.25\times\ln0.25=\ln4\approx1.39$.\\

The sum of all one-orbital entropies gives a measure for the \emph{total correlation}
\begin{equation}
    I_{\mathrm{tot}} = \sum_i S_i
\end{equation}
of the wave function \cite{Legeza-2004b,Szalay-2017}.\\

Analogues to the one-orbital von Neumann entropy, higher orders can be calculated from their corresponding $n$-orbital reduced density matrix. For example the \emph{two-orbital} von Neumann entropy $S_{ij}$ is obtained from the \emph{two-orbital} density matrix $\rho_{ij}$ \cite{Legeza-2006a}
\begin{align}
    \rho_{ij}(\alpha_i,\alpha_j,\alpha_i^\prime,\alpha_j^\prime) &= \mathrm{Tr}_{1,\dots,\not{i},\dots,\not{j},\dots,d}|\Psi\rangle\langle\Psi| \\
    S_{ij} &= -\mathrm{Tr}\left(\rho_{ij}\ln\rho_{ij}\right) = -\sum_\alpha \omega_{ij,\alpha} \ln \omega_{ij,\alpha}
\end{align}
and quantifies the correlation contributions of the combined two-orbital subsystem $ij$. To quantify correlations between $i$ and $j$ the mutual information $I_{ij}$ \cite{Rissler-2006} has to be calculated
\begin{equation}
    I_{ij} = S_i + S_j - S_{ij}.
    \label{eq:I_ij}
\end{equation}

\subsection{Comparison of Increments and Entropies}
Both, MoI and QIT, allow for quantification of correlation contributions by single orbitals or groups of them. If we assign each orbital to its own center in the MoI, we can directly compare the increments ($\varepsilon_i$, $\Delta\varepsilon_{ij}$) with the QIT quantities ($S_i$, $I_{ij}$). To stress this assignment, we will switch terminology from here on, and call the one- and two-\emph{center} increments, one- and two-\emph{orbital} increments respectively.\\

Indeed a close similarity is obvious when comparing \cref{eq:2center} and \cref{eq:I_ij}. The two equations suggest a close connection of the two-orbital increment $\Delta\varepsilon_{ij}$ with the mutual information $I_{ij}$, as well as of the one-orbital increments $\varepsilon_i$ with the one-orbital entropy $S_i$ and the two-orbital correlation energy $\varepsilon_{ij}$ with the two-orbital entropy $S_{ij}$.
Therefore, one might ask whether it is possible to infer from the QIT quantities to the increments and vice versa? \\

It should be pointed out however, that there are some main differences between entropies and increments. First of all, in the MoI one only accounts for correlation effects of the current increment with the rest of the system. For example when calculating the two-orbital increments $\varepsilon_{ij}$ the correlations effects of all other orbitals $k\neq i\neq j$ are not considered, i.e.~each increment is only correlated with the virtual orbitals, but not the other increments on its level. On the other hand, the QIT entropies rely on a CI wave function which correlates all orbitals at once.\\

The second main difference is the definition of both quantifies. The increments are essentially energies obtained by applying the Hamiltonian to different truncations of the CI wave function, while for the entropies we first construct reduced density matrices (all from the same CI wave function), and then apply a logarithmic function (cf. \cref{eq:S1}) to its eigenvalues, making it a non-linear mapping.\\

Thus there is no direct correspondence of increments and entropies, and quantitative differences are to be expected. We will show however, that qualitative agreement can indeed be observed.\\

\section{Model Systems}
\label{sec:models}

We will investigate both measures, increments an entropies, in two different model system. We need to restrict ourselves to closed shell systems, as application of the MoI to open shell systems is currently limited to systems with only a single unpaired electron \cite{Mueller-2012}.\\

\subsection{Polyacetelene}
For the first basic molecular model system we choose conjugated \emph{trans}-polyacetelenes, whose single and double bonds will provide varying degrees of correlation effects. MoI calculations have been successfully applied to such systems before, and allowed to discriminate between different bonding situations \cite{Yu-1997}.
Additionally, linear systems are better suited for the linear MPS structure in DMRG calculations. For our study we choose the hexatriene (\ce{C6H8}), which is just large enough to allow for spatially well separated localized orbitals.
Furthermore it provides two different kinds of double bonds, one at the center of the chain, the other two on each end of the chain. The used bond distances are \SI{145}{\pico\meter} for \ce{C-C} bonds, \SI{136}{\pico\meter} for \ce{C=C} bonds and \SI{109}{\pico\meter} for \ce{C-H} bonds, while all bond angles are set to \SI{120}{\degree}.\\

Hartree-Fock orbitals are obtained using the cc-pVTZ basis set \cite{Kendall-1992} and occupied orbitals are localized using Pipek-Mezey \cite{Pipek-1989} localization. For the active space we exclude the 6 $1s$ carbon core orbitals, and include all orbitals which can be constructed from the carbon $2sp$ and hydrogen $1s$ shells, resulting in 16 occupied and 16 virtual orbitals. We thus use for both, DMRG and MoI, an active space with 32 electrons in 32 orbitals (CAS(32,32)). \\

\subsection{Beryllium Ring}

As a second system we will study the \ce{Be6} ring in $D_{6h}$ symmetry at equilibrium distance ($R=\SI{2.2}{\angstrom}$) and the dissociation limit ($R=\SI{3.5}{\angstrom}$). This system has been studied before by means of DMRG and QIT to model the metal-insulator-transition \cite{Fertitta-2014}. Another study applied MoI and investigated the effect of different basis sets and correlation methods the MoI is based on \cite{Koch-2016}.\\

For the \ce{Be6} ring we first calculate the Hartree-Fock orbitals in the cc-pVDZ basis set \cite{Kendall-1992}.
Note that for the equilibrium and dissociation regime different Hartree-Fock configurations need to be considered \cite{Fertitta-2014}.
All occupied and virtual molecular orbitals are then localized by the Foster-Boys scheme \cite{Boys-1960}.\\

Due to the localization all orbitals are grouped in sets of 6 degenerate orbitals, which can be transformed into each other by $C_6$ rotations perpendicular to the molecular plane.
This degeneracy results in a strong correlation, especially for the dissociated situation.\\

The six $1s$ core orbitals are regarded as closed for all active space considerations, which leaves another six occupied HF orbitals with 12 electrons for the active space.
Choosing the virtual orbitals for the active space is more difficult this time.
As the virtual orbitals do not contribute to the Hartree-Fock ground state energy, there is no driving force which might lead to meaningful orbitals with any physical or chemical interpretation.
We therefore first calculate all virtual 1-orbital increments (cf. \cref{sec:compdetails}) and then remove virtual orbitals from the active space based on these values for DMRG calculations.
This is a computationally very cheap task as each 1-orbitals increments corresponds to one (12,7) active space calculation, i.e.~there is only a single empty orbital.
However, selecting virtual orbitals based on QIT results is not feasible, because the whole virtual space involves dynamical correlation effects, which DMRG is not designed for. Instead we construct the DMRG active space by only considering virtual orbitals above a user-defined 1-orbital increment threshold.\\

\section{Computational Details}
\label{sec:compdetails}

For the DMRG calculations the Budapest DMRG program \cite{DMRGcode} was used, while all other calculations were performed using \textsc{Molpro} \cite{MOLPRO, MOLPRO-WIREs}.
For the DMRG and QIT results of polyacetelene presented in \cref{tab:Eref} and \cref{fig:MoIvsQIT} the Dynamic Block State Selection (DBSS) approach\cite{Legeza-2003a,Legeza-2004b} was used, with a density matrix cutoff of $1\times10^{-6}$ and a maximum number of block states of $M_{\rm max}=2048$. Together with an optimized orbital ordering these parameters provide high accuracy for the correlation energy. For the QIT quantities however, it is sufficient to use $M_{\rm max}=128$ provided an appropriate orbital ordering is used. For the \ce{Be6} ring we have used a density matrix cutoff of $1\times10^{-5}$ and a maximum number of block states of $M_{\rm max}=1024$, as the larger number of orbitals requires more memory.\\

The QIT results will give us entanglement measures among occupied and virtual orbitals, as well as in between them in just one run. However, in the conventional MoI approach, we can only get the correlations for (groups of) occupied centers. In order to compare both measures (QIT and MoI) directly we define each occupied orbital to be its own center. Furthermore, to access the correlation measures for virtual orbitals, we can flip things around and expand the increments in virtual orbitals and correlate these centers with all occupied orbitals. For large virtual spaces, this will drastically reduce the active space for each individual increment calculation and in turn increase the number of centers. This approach has recently also been suggested by \textcite{Eriksen-2017} in an effort to improve parallelism of such calculations and make large orbital spaces more accessible.\\

\section{Results}
\label{sec:results}

\begin{table}
    \centering
    \caption{Total energy and correlation energy for \emph{trans}-hexatriene obtained with different methods using a cc-pVTZ basis set. All energies are in \si{\hartree}.}
    \label{tab:Eref}
    \begin{ruledtabular}
    \begin{tabular}{ccc}
        & Total Energy & Correlation Energy \\
        \toprule
        HF          & $-231.883732$ & \\
        DMRG(32,32) & $-231.971288$ & $-0.087556$ \\
        CCSD        & $-232.891987$ & $-1.008255$ \\
    \end{tabular}
    \end{ruledtabular}
\end{table}

\begin{table}
    \centering
    \caption{Correlation energies obtained with various MoI variants for \emph{trans}-hexatriene using a cc-pVTZ basis set. Occupied orbitals have been localized using the Pipek-Mezey method. All energies are in \si{\hartree}. }
    \label{tab:MoI}
    \begin{ruledtabular}
    \begin{tabular}{lccc}
        & Level & Correlation Energy & Summed Correlation Energy \\
        \toprule
        CAS(32,32)-MoI (occupied)   & 1 & $-0.053475$ & $-0.053475$ ($\hphantom{1}\SI{61.1}{\%}$)\textsuperscript{a} \\
                                    & 2 & $-0.035636$ & $-0.089111$ ($\SI{101.8}{\%}$)\textsuperscript{a} \\
                                    & 3 & $+0.000988$ & $-0.088122$ ($\SI{100.6}{\%}$)\textsuperscript{a} \\
                                    & 4 & $+0.000393$ & $-0.087729$ ($\SI{100.2}{\%}$)\textsuperscript{a} \\
        \hline
        CAS(32,32)-MoI (virtual)    & 1 & $-0.022884$ & $-0.022884$ ($\hphantom{1}\SI{26.1}{\%}$)\textsuperscript{a} \\
                                    & 2 & $-0.063367$ & $-0.086251$ ($\hphantom{1}\SI{98.5}{\%}$)\textsuperscript{a} \\
                                    & 3 & $-0.003104$ & $-0.089355$ ($\SI{102.1}{\%}$)\textsuperscript{a}\\
                                    & 4 & $+0.001568$ & $-0.087787$ ($\SI{100.3}{\%}$)\textsuperscript{a}\\
        \hline
        CCSD-MoI (occupied)         & 1 & $-0.511070$ & $-0.511070$ ($\hphantom{1}\SI{50.7}{\%}$)\textsuperscript{b} \\
                                    & 2 & $-0.552483$ & $-1.063553$ ($\SI{105.5}{\%}$)\textsuperscript{b} \\
                                    & 3 & $+0.060588$ & $-1.002969$ ($\hphantom{1}\SI{99.5}{\%}$)\textsuperscript{b}\\
    \end{tabular}
    \end{ruledtabular}
    \flushleft
    \vspace{0.1\baselineskip}
    a) Correlation energy relative to DMRG(32,32) reference\\
    b) Correlation energy relative to CCSD reference.
\end{table}

We start our discussion with the total energies calculated for the polyacetelene system by different methods. The energies in \cref{tab:Eref} provide references for the uncorrelated system (HF), the static correlation contribution (DMRG) and dynamic correlations (CCSD). In \cref{tab:MoI} we present the MoI results at different levels of increments, again dealing with static (CAS(32,32)-MoI) and dynamic correlation (CCSD-MoI). The 3-orbital increments level yields in all cases agreement within $10^{-2} E_{\mathrm{h}}$, while with 4-orbital increments reduce the difference by one order of magnitude to $10^{-3} E_{\mathrm{h}}$. In case of the CAS-MoI, we also expanded the correlation energy in terms of virtual orbitals, yielding similar results when including the 3- and 4-orbital increments.

\subsection{Polyacetelene: Static Correlation}

\begin{figure}
    \centering
    \subfloat[][\#12]{\includegraphics[width=0.32\textwidth]{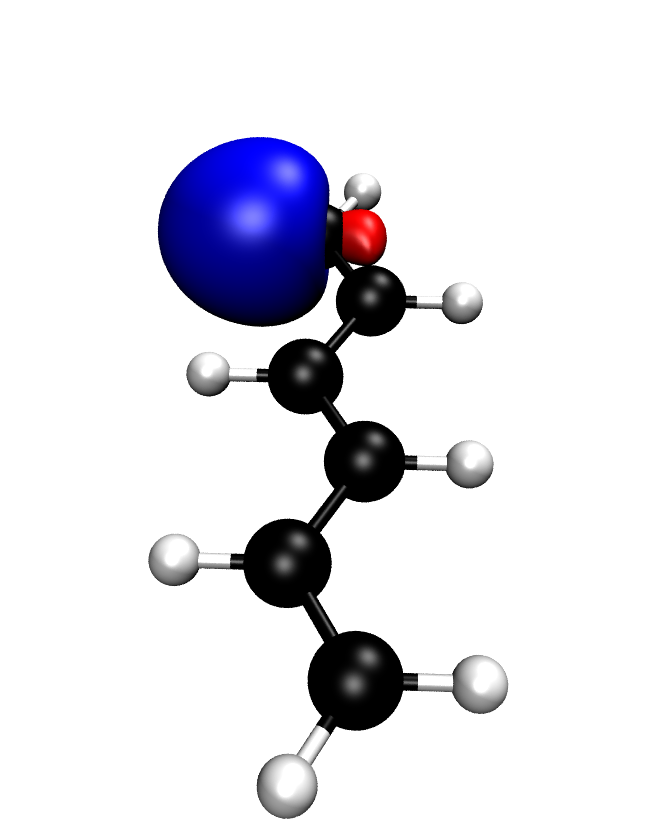}}
    \subfloat[][\#13]{\includegraphics[width=0.32\textwidth]{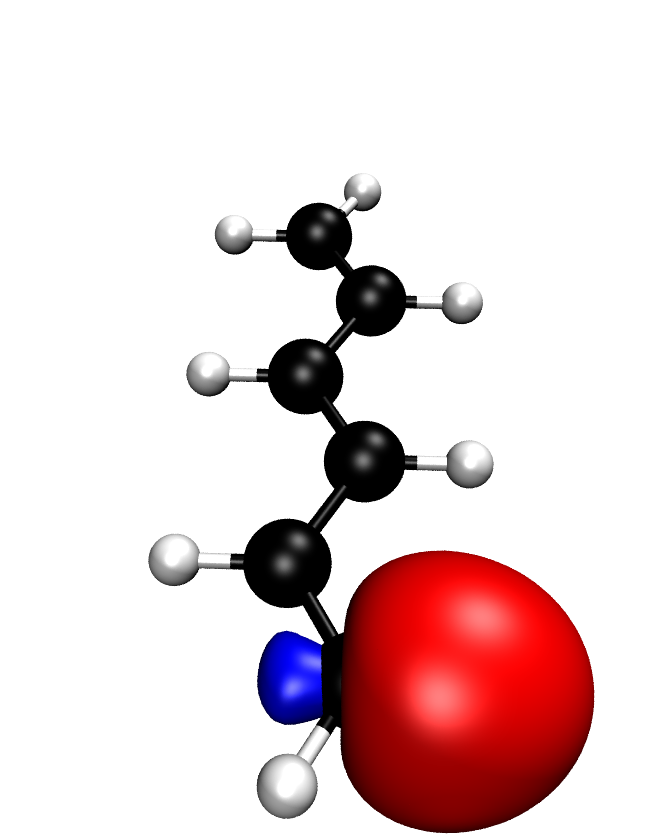}}
    \subfloat[][\#14]{\includegraphics[width=0.32\textwidth]{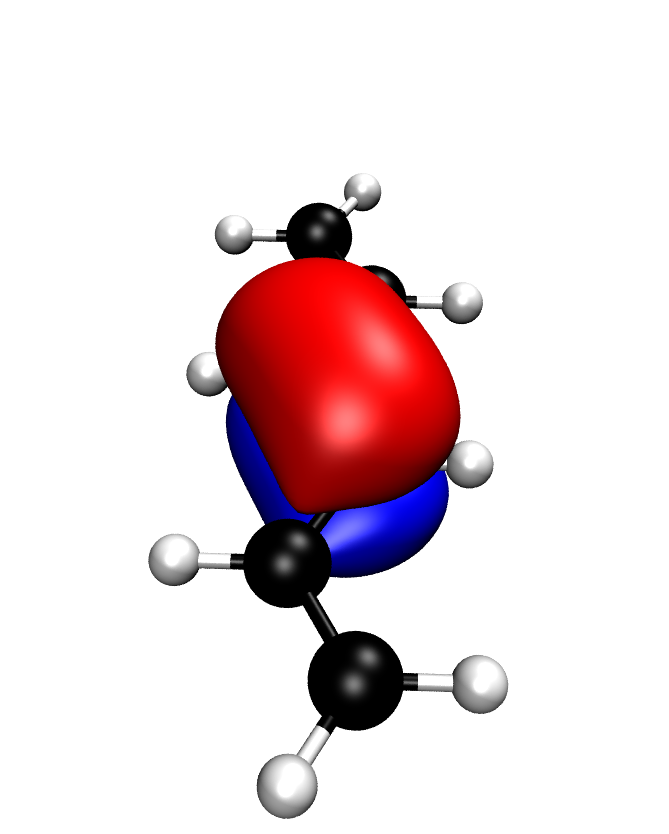}}\\
    \subfloat[][\#15]{\includegraphics[width=0.32\textwidth]{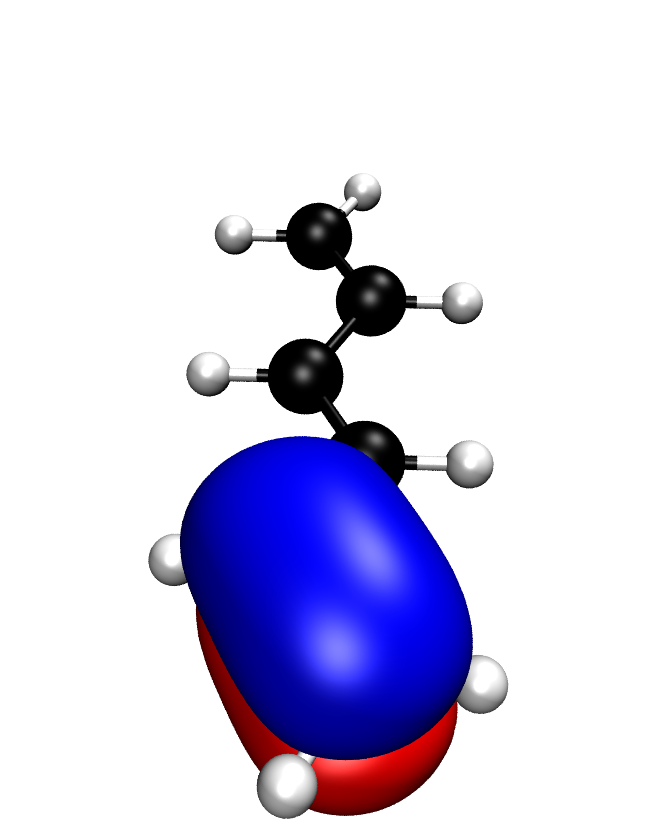}}
    \subfloat[][\#16/HOMO]{\includegraphics[width=0.32\textwidth]{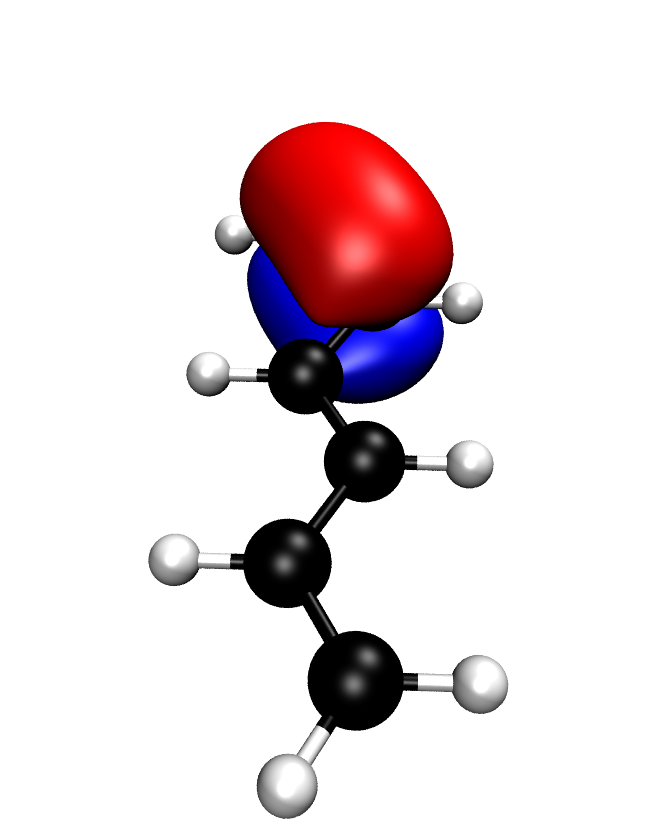}}
    \subfloat[][\#17/LUMO]{\includegraphics[width=0.32\textwidth]{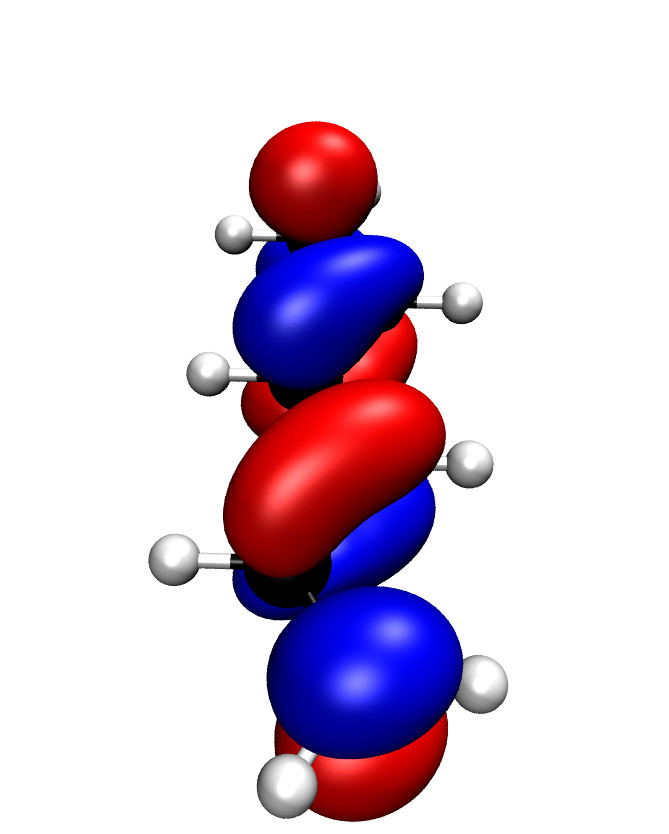}}
    \caption{Isosurface plot ($\psi(\vec r)=0.05\ a_0^{-1.5}$) for the localized occupied molecular orbitals \#12 to \#16 and the canonical virtual molecular orbital \#17.}
    \label{fig:MOs}
\end{figure}

\begin{figure}
    \centering
    \includegraphics[width=0.49\textwidth]{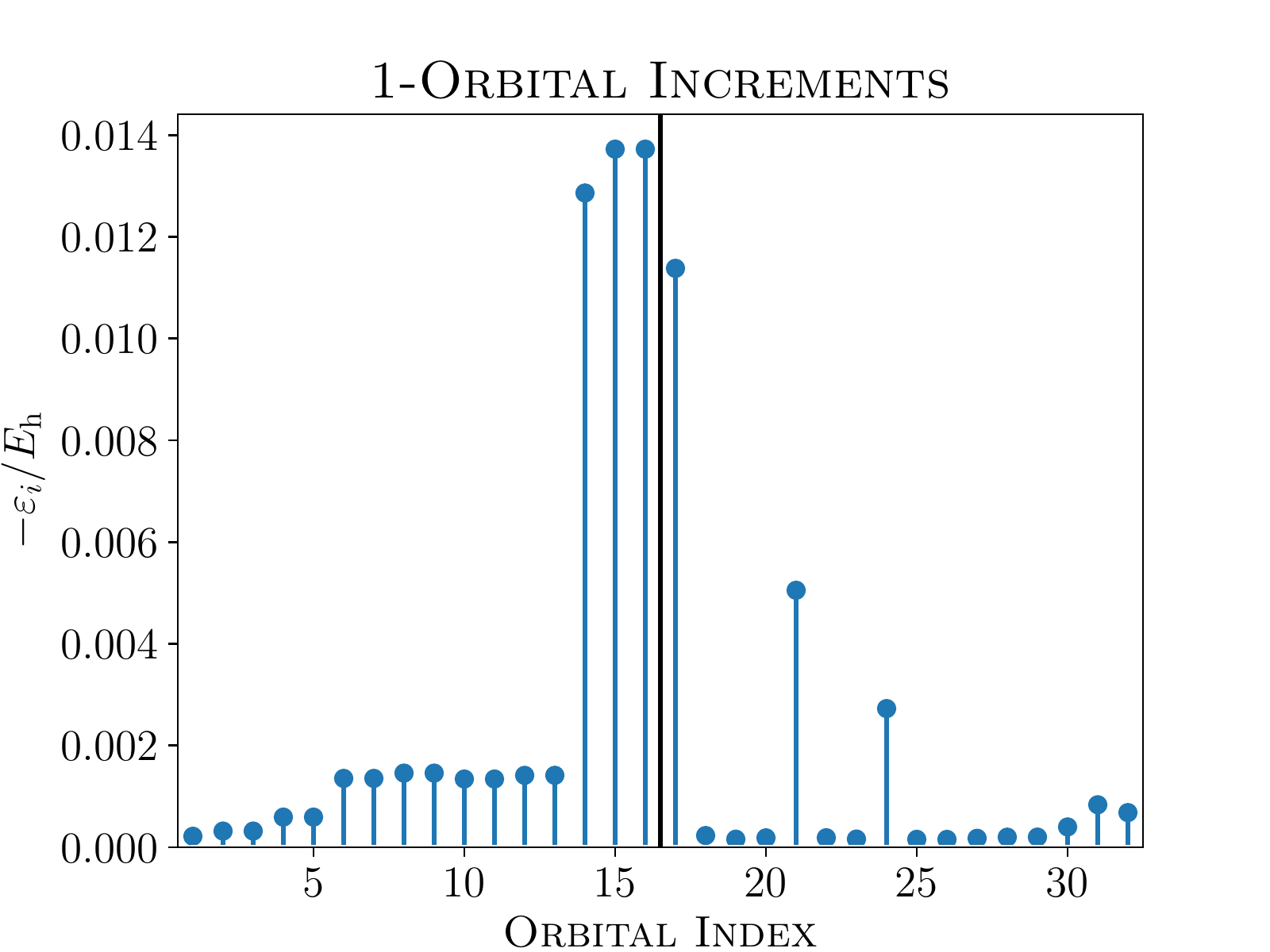}
    \includegraphics[width=0.49\textwidth]{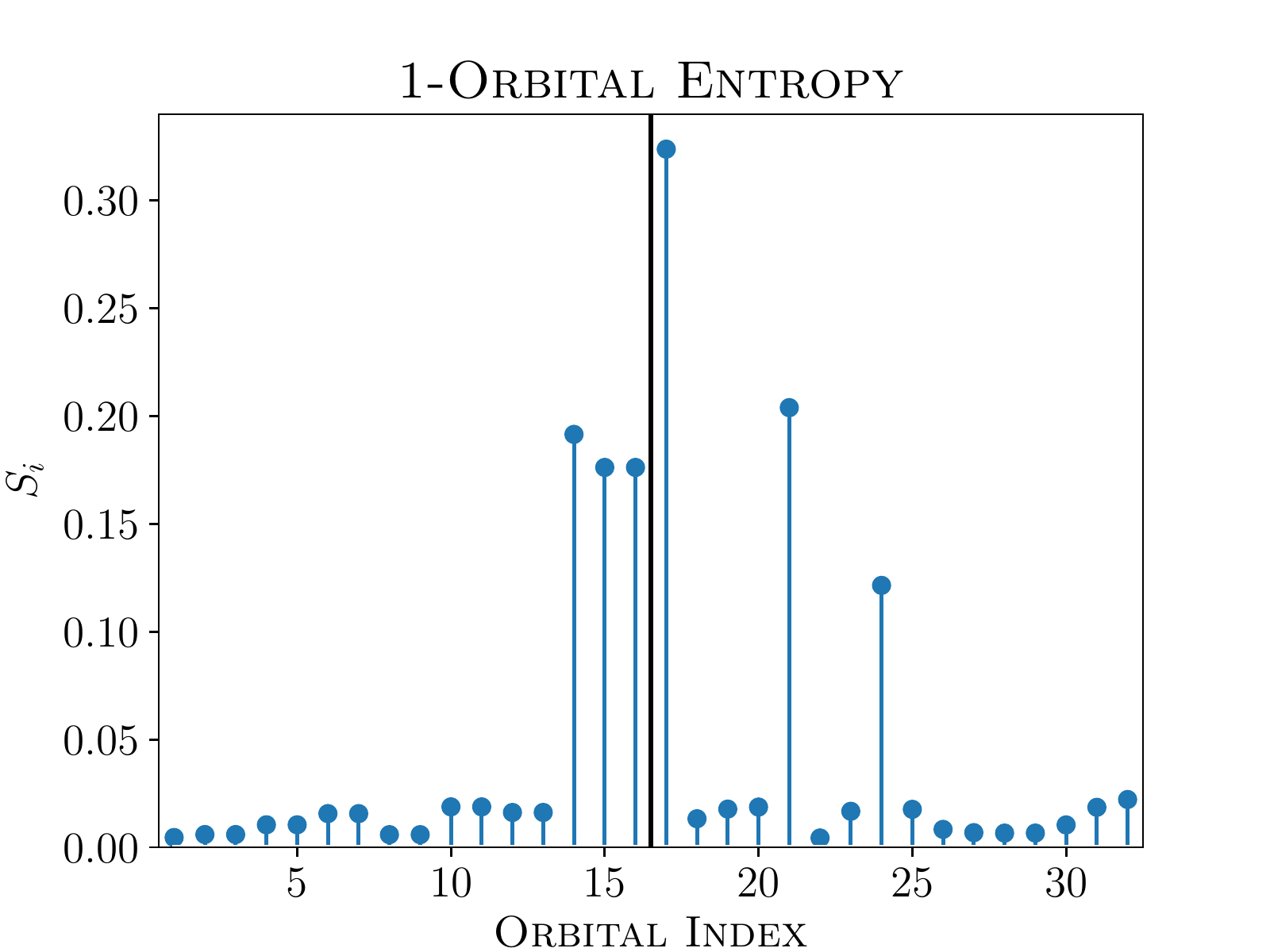}\\
    \includegraphics[width=0.49\textwidth]{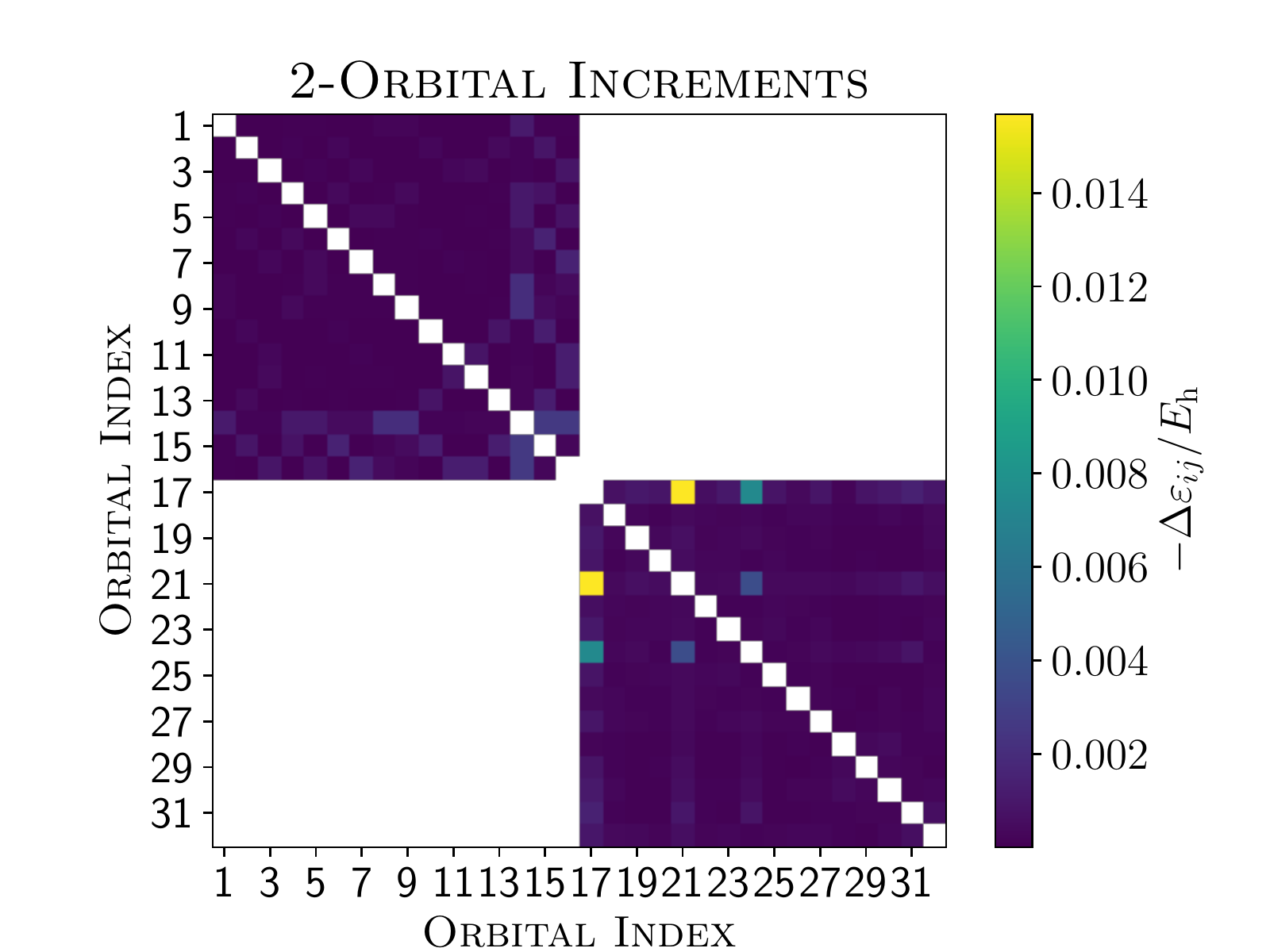}
    \includegraphics[width=0.49\textwidth]{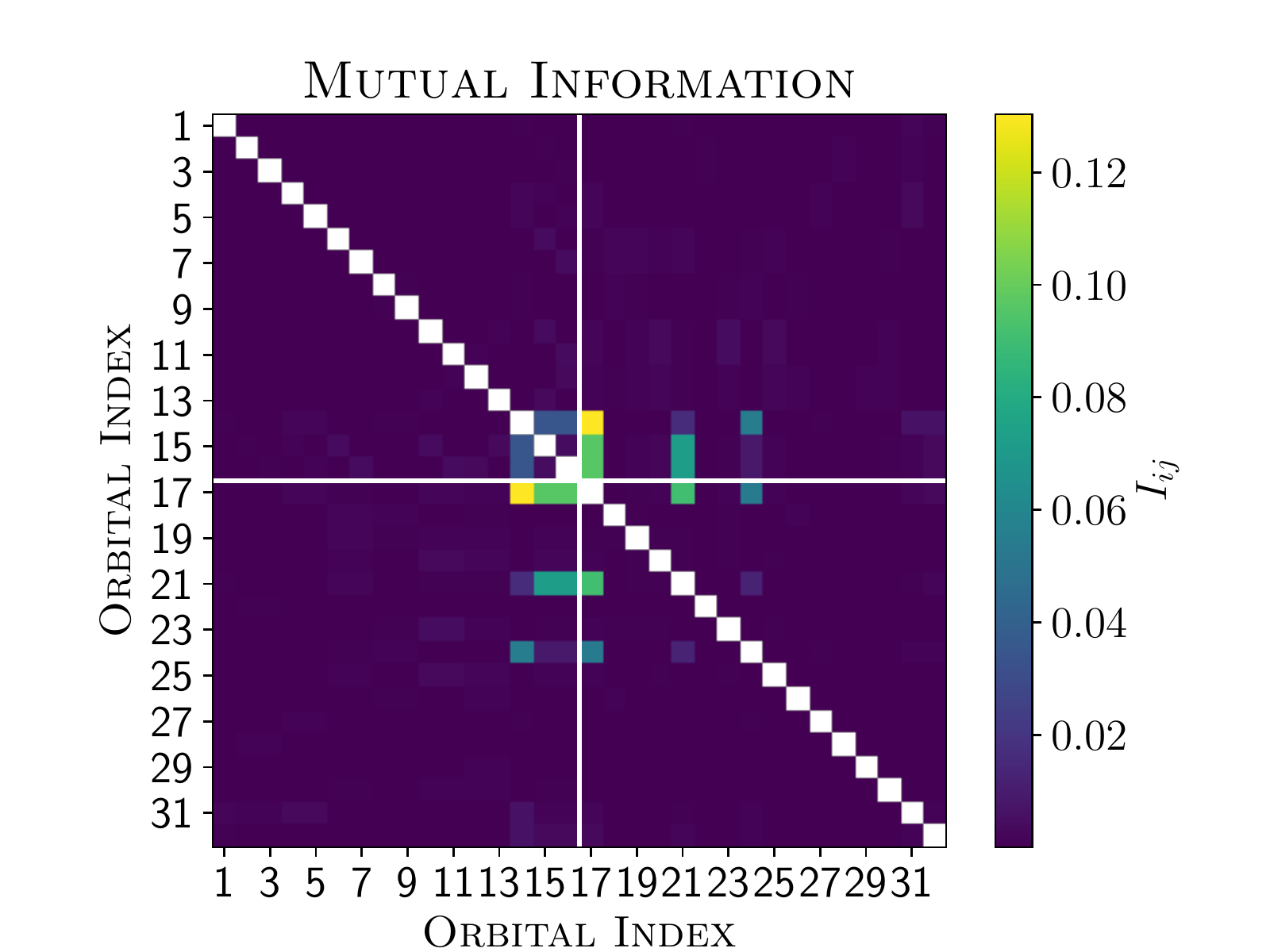}\\
    \includegraphics[width=0.49\textwidth]{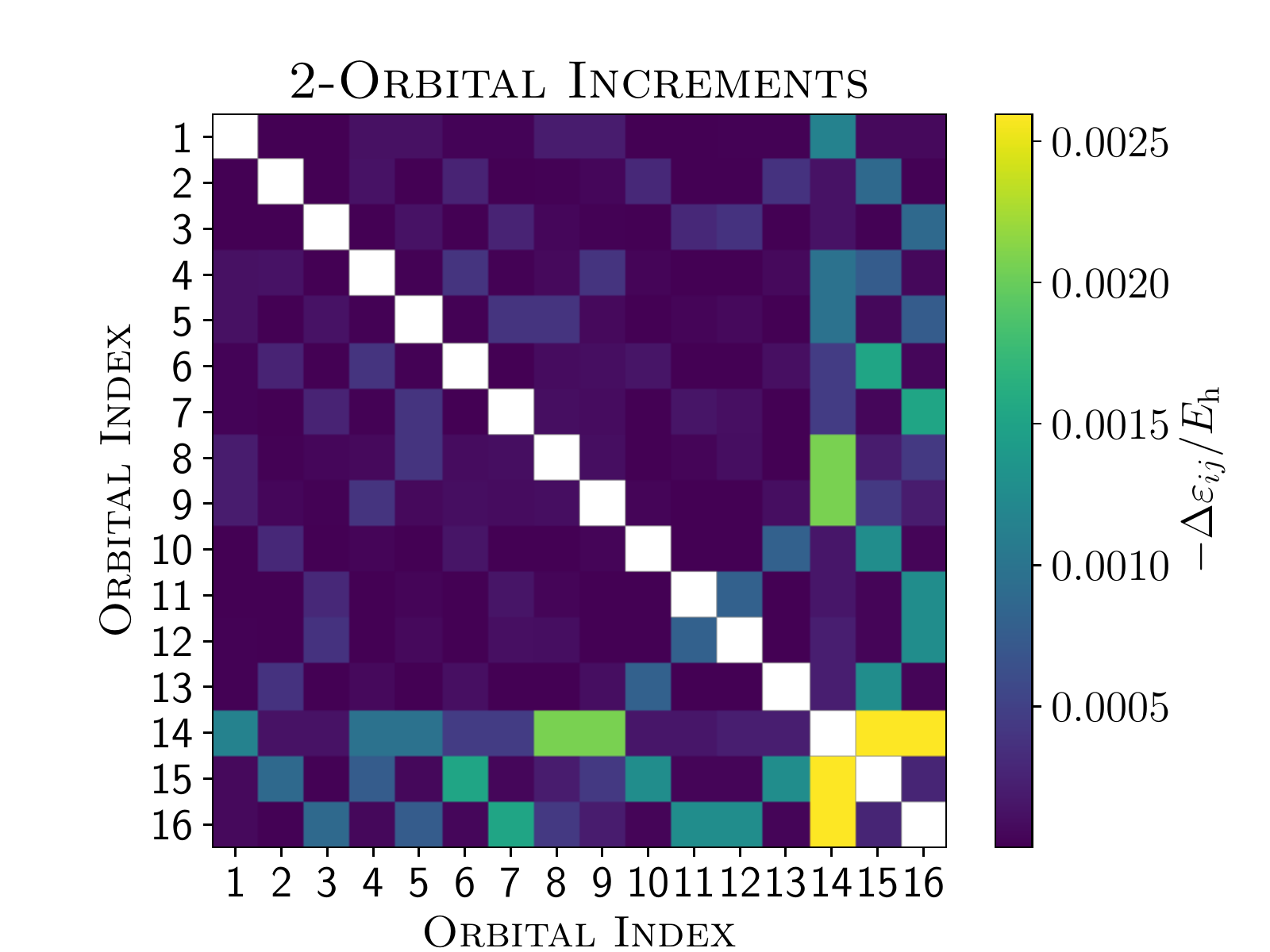}
    \includegraphics[width=0.49\textwidth]{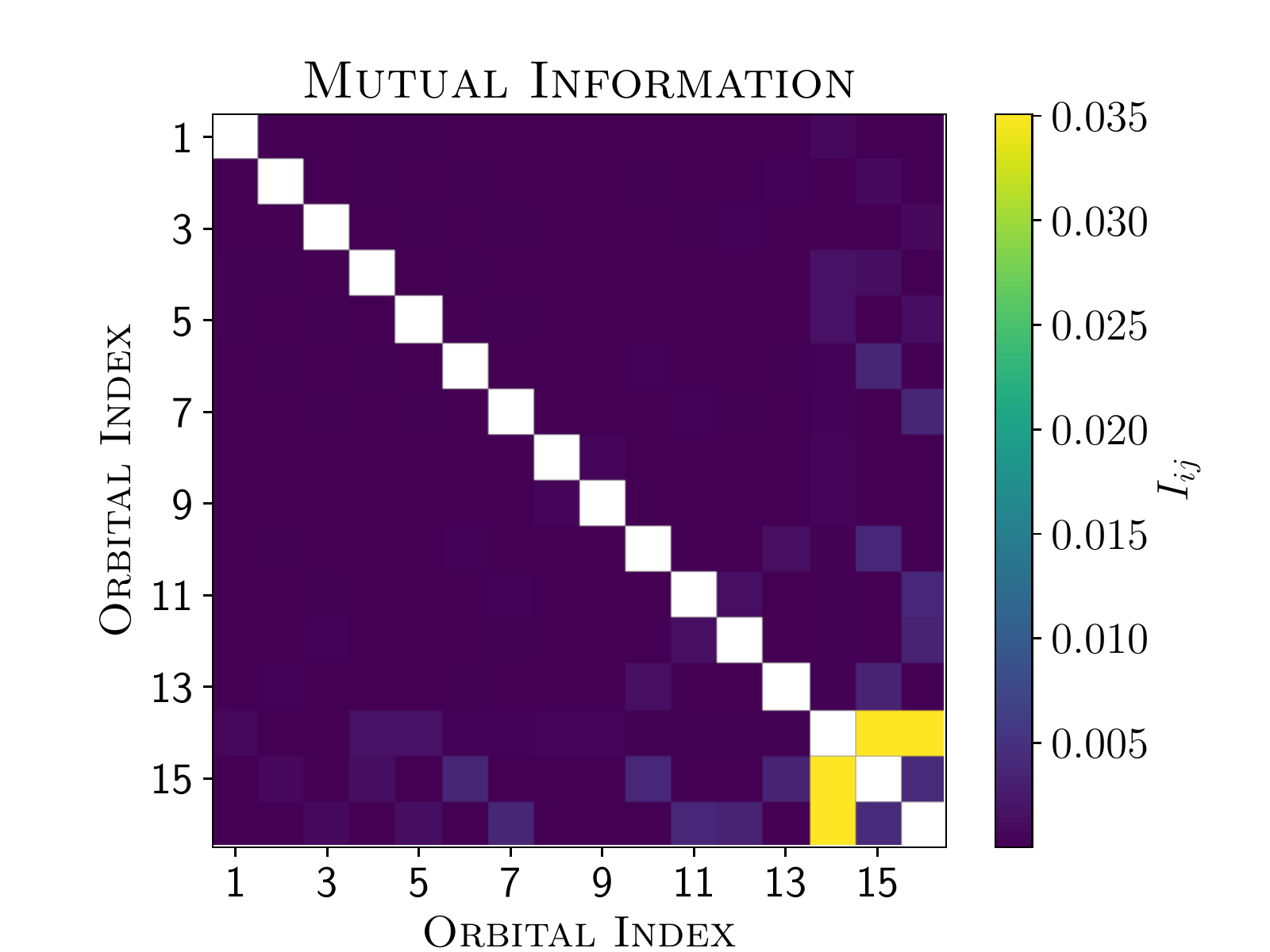}
    \caption{Polyacetelene: Comparison of Increments and Orbital Entropies for quantifying orbitals correlations. Left and right column show results based on Methods of Increments (MoI) and Quantum Information Theory (QIT) respectively. The upper and middle row show 1-orbital and 2-orbital correlations respectively. The Highest Occupied Molecular Orbital (HOMO) and Lowest Unoccupied Molecular Orbital (LUMO) are separated by black and white lines respectively. In the lower row the 2-orbital correlations among occupied orbitals are shown as a zoom in. MoI increments are based on CASCI(32,32)/cc-pVTZ calculations.}
    \label{fig:MoIvsQIT}
\end{figure}

Before investigating the increments and entropies we take a brief look at a selected set of the molecular orbitals \cref{fig:MOs}. For a complete list of the active space orbitals please refer to the Supplementary Information (Figs.~S1 to S4).
We will focus our discussion to the exemplary two-orbital correlations involving localized occupied molecular orbitals describing \ce{C-H} $\sigma$ bonds (\#12 and \#13) and \ce{C=C} $\pi$ bonds (\#14 to \#16) together with the Lowest Unoccupied Molecular Orbital (LUMO) \#17. The remaining occupied orbitals describe the \ce{C-C} $\sigma$ bonds (\#1 to \#5) and further \ce{C-H} $\sigma$ bonds (\#6 to \#11).
Note that \ce{C-H} orbitals \#12 and \#13 as well as  \ce{C=C} $\pi$ orbitals \#15 and \#16 are localized at the ends of the linear molecule, which results in examples for spatially close lying and well separated two-orbital correlations.\\

In \cref{fig:MoIvsQIT} we present the one- and two-orbital increments (left column) as well as one-orbital entropy and mutual information (right column). Note that the two-orbital increments are calculated by doing two separate incremental expansions: the first one in terms of occupied orbitals, the second one in terms of virtual orbitals. The results are however compiled in a single picture. This cut between occupied and virtual orbitals is indicated by black and white lines for the one- and two-orbital quantifies respectively. The off-diagonal occupied-virtual pairs for the two-orbital increments cannot be calculated by the MoI method, therefore these missing values are indicated in white. \\

Comparing both columns we find very similar patterns, but with varying amplitudes. It is evident from both the one-orbital increments and one-orbital entropy (upper row in \cref{fig:MoIvsQIT}), that the orbitals with largest correlation contributions are \#14 to \#17, representing the $\pi$ bonds. Out of these four, the one-orbital increments identify orbitals \#15 and \#16, the $\pi$ bond on each end of the chain, to be the most important one, and the LUMO \#17 to be the least important one. However, the 1-orbital entropies shows a reversed trend, and even includes the virtual orbital \#21 as the second highest one.
Both approaches agree though in the trend for the $\sigma$ bond orbitals: orbitals \#1 to \#5 (\ce{C-C}) are smaller than orbitals \#6 to \#13 (\ce{C-H}), being much smaller than the $\pi$ orbitals.\\

The two-orbital correlations (middle row in \cref{fig:MoIvsQIT}) exhibit similar patterns as well, again with varying amplitudes. Note however, that the largest contributions identified by the mutual information, namely orbital pairs \#17\#14, \#17\#15 and \#17\#16, are not included in the MoI as they include one occupied and one virtual orbital. Focusing on the occupied correlations only (bottom row in \cref{fig:MoIvsQIT}), we see that both methods again agree in the general trend: most important correlation contributions arise from the three bonding $\pi$ orbitals, while the spatially separated pair \#15\#16 is negligible. All the degenerate pairs of $\sigma$ bond orbitals (e.g. \#12 and \#13) show identical correlation contributions when paired with the centered $\pi$ orbital \#14, as their distance is the same. In connection with the $\pi$ orbitals on the ends of the chain (\#15, \#16) an alternating structure of large and small correlations emerges from the varying spatial separation.\\

\subsection{Polyacetelene: Dynamic Correlation}
The inclusion of dynamical correlations by increasing the virtual orbital space leads costly calculations for the DMRG based QIT results. The MoI however, can easily be based on the Coupled Cluster Singles Doubles (CCSD) method.
The latter will be identical to Full Configuration Interaction (FCI) calculations for 1-orbital increments, as only two electrons are correlated in each individual calculation. Similar, for the 2-orbital increments with 4 electrons each, we can expect the CCSD error to be negligible.
Furthermore, increments expanded in virtual orbitals will be the same as for the CAS-MoI results above, as the occupied orbital space remains unchanged. It will however add contributions of higher lying virtual orbitals.
\\

\begin{figure}
    \centering
    \includegraphics[width=0.49\textwidth]{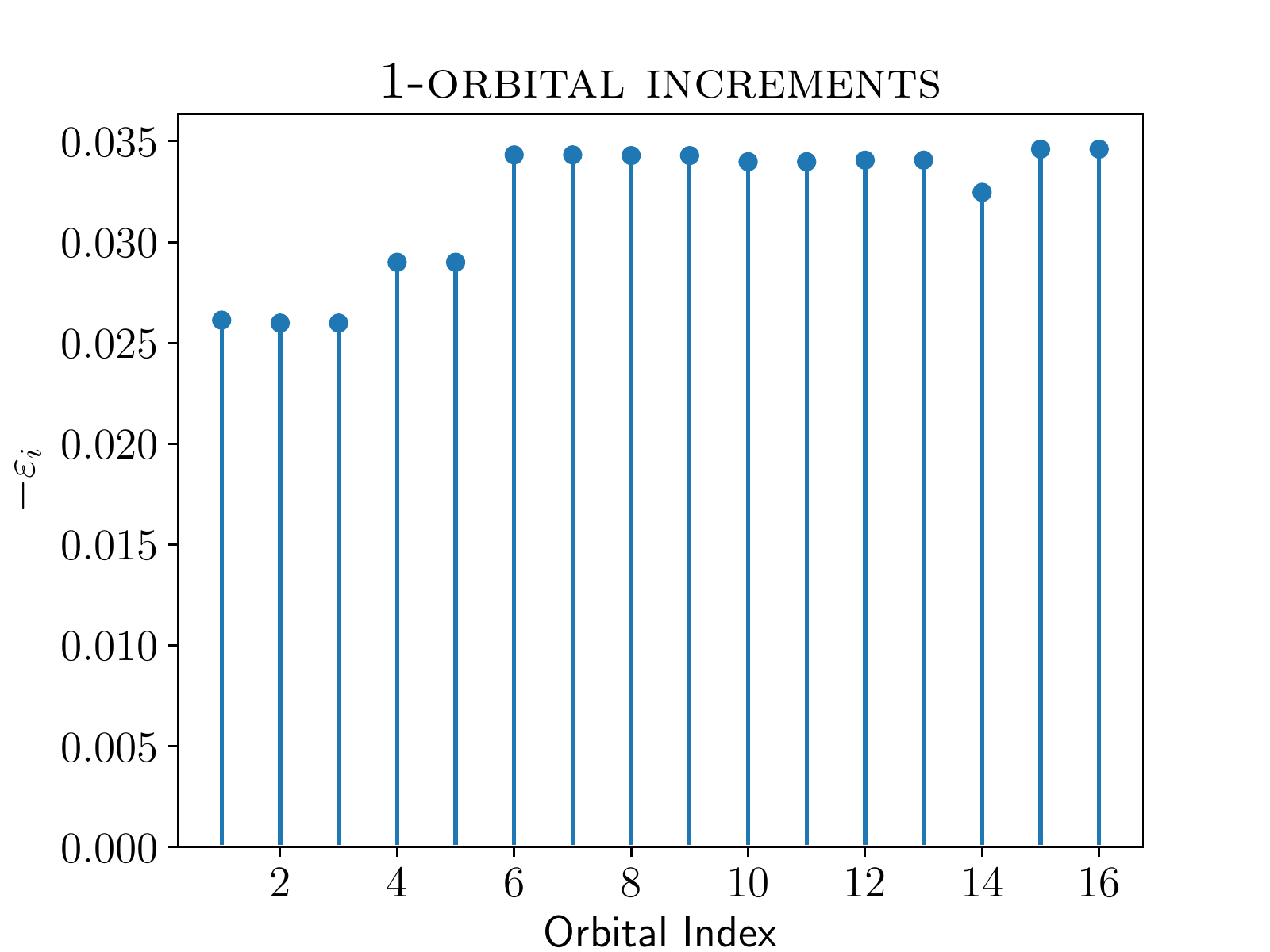}
    \includegraphics[width=0.49\textwidth]{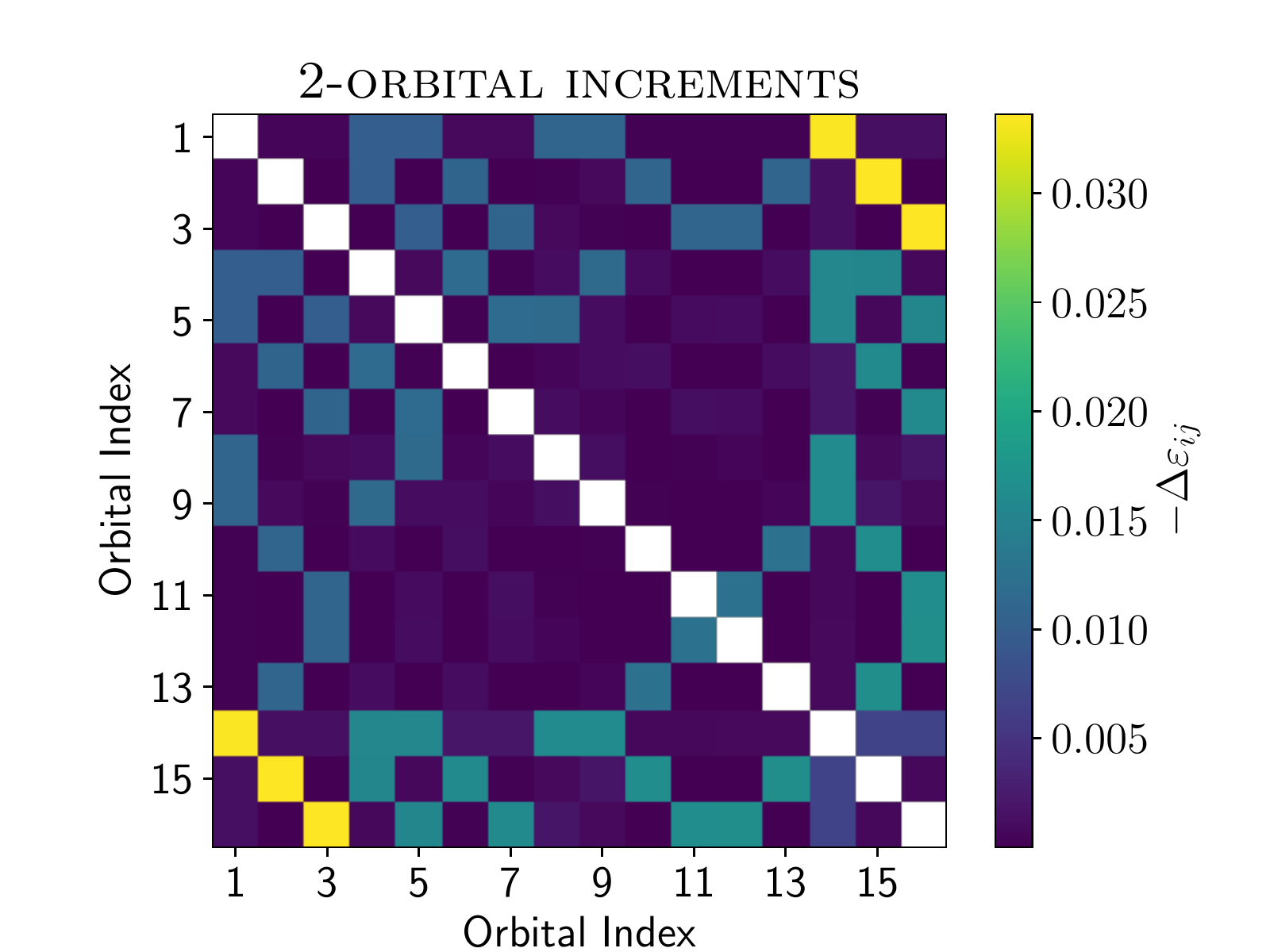}
    \caption{Polyacetelene: Occupied orbital increments including dynamical correlations based on CCSD/cc-pVTZ calculations.}
    \label{fig:CCSD}
\end{figure}

CCSD-MoI results expanded in terms of occupied orbitals are shown in \cref{fig:CCSD}. We observe a large energy shift for all increments, as the virtual orbital space increased drastically. This indicates dynamical correlations being of similar importance for all occupied orbitals. However, the effect is less pronounced for the $\pi$ orbitals (\#14 to \#16), resulting in similar values as for the \ce{C-H} bond orbitals (\#6 to \#12). Only the \ce{C-C} sigma bond orbitals (\#1 to \#5) remain distinguishable by their value.\\

The 2-orbitals increments exhibit a very similar pattern, but quite large differences in the magnitude.
Although being the smallest 1-orbital increments, the orbitals \#1 to \#3 now have largest contributions in the 2-orbital increments. These 2-orbital increments correspond to the combinations of \ce{C-C} $\sigma$ and $\pi$ bond, being localized on the same bonds (\#1\#14 representing the central \ce{C-C} bond; \#2\#15 and \#3\#16 the two outer most \ce{C-C} bonds; cf. Supplementary Information).
We can therefore conclude, that smaller lower-level increments do not necessarily indicated negligible higher-level contributions, thus care must be taken when trying to deduce from one to the other.
Furthermore we can observe that the 2-orbital increments for the two neighboring $\pi-\pi$ pairs is now one of the smallest contribution (but still with increased value compared to the CAS-MoI results).
The colors for \ce{C-C} $\sigma-\pi$ and \ce{C-C} $\pi-\pi$ are essentially reversed for static and dynamical correlation effects. Thus we can see that correlation effects between spatially overlapping $\sigma$ and $\pi$ bond are mainly of dynamical nature, in contrast to the static correlations between neighboring, degenerate $\pi$ orbitals.\\

\subsection{Beryllium Ring}

Plots and numerical data for all active space orbitals are available in the Supplementary Information (Figs.~S7 to S12 for $R=\SI{2.2}{\angstrom}$ and Figs.~S13 to S19 for $R=\SI{3.5}{\angstrom}$).
After applying the Foster-Boys localization most of the molecular orbitals are localized on a single \ce{Be} atom with a 6-fold degeneracy. However, in case of the equilibrium distance ($R=\SI{2.2}{\angstrom}$), there are some exceptions to that. The six bonding molecular orbitals together with a set of six anti-bonding are localized between the \ce{Be}, i.e.~on the bond. Those are similar to the orbitals already reported by \textcite{Fertitta-2014}, determined with a smaller basis set.
Regarding the localization of the virtual orbitals at equilibrium distance, we had some numerical issues. Due to the $D_{6h}$ symmetry of the system, the diagonal Fock-Matrix elements should be 6-fold degenerate as well, but they only agree up to \SI{0.01}{\hartree} (cf. Figs.~S7 to S12 of the Supplementary Information). However, the difference is not visible in the plotted iso-surfaces. We can therefore expect some minor deviations in the entropies and increments as well.\\

\begin{figure}
    \centering
    \includegraphics[width=0.49\textwidth]{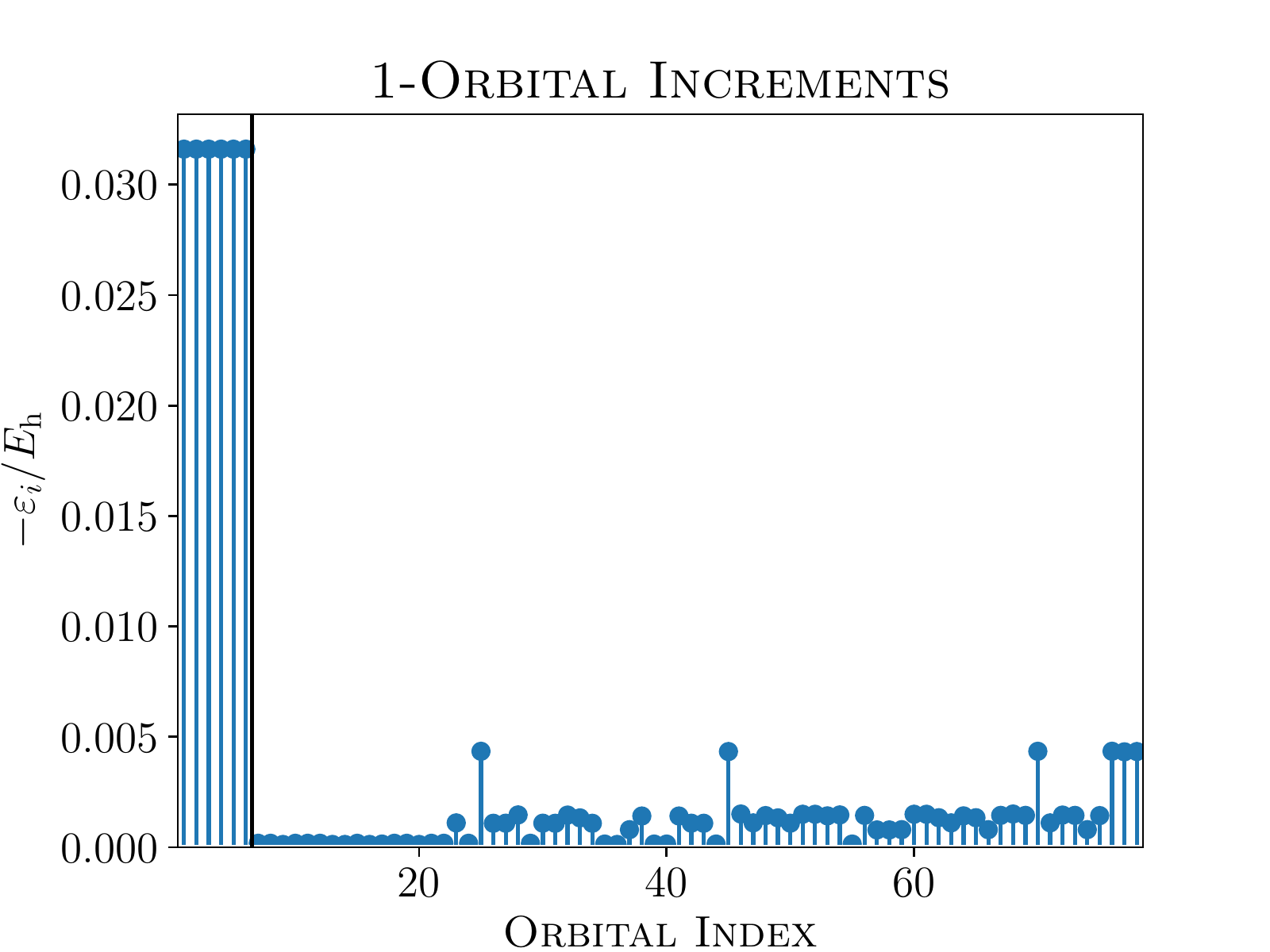}
    \includegraphics[width=0.49\textwidth]{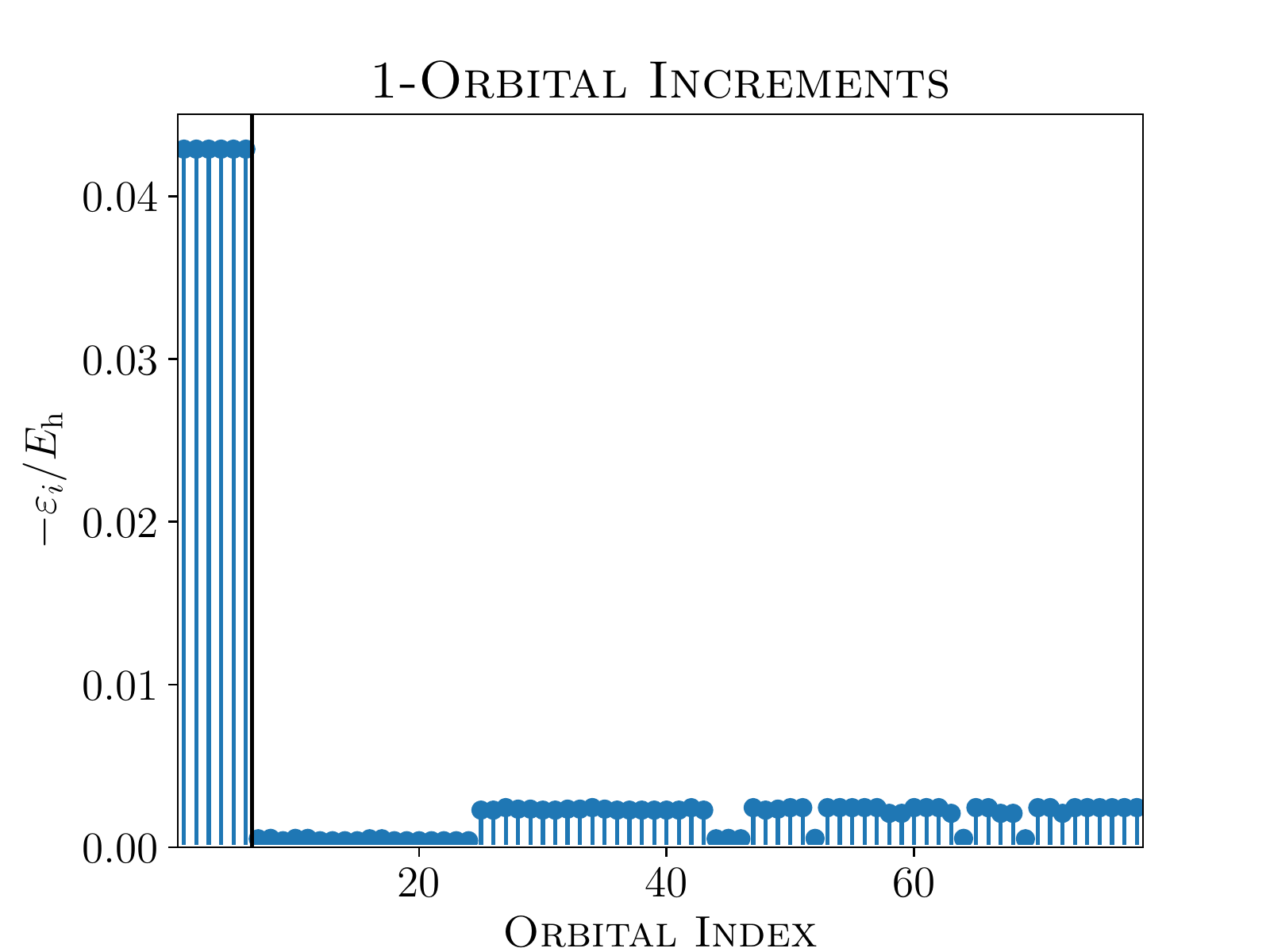}\\
    \caption{1-Orbital increments for the \ce{Be6} ring at equilibrium distance ($R=\SI{2.2}{\angstrom}$) and dissociation limit ($R=\SI{3.5}{\angstrom}$) to the left and right respectively. The black vertical line separates occupied and virtual orbitals. Virtual orbitals are in arbitrary order.}
    \label{fig:Be6_allvirtual}
\end{figure}

\begin{figure}
    \centering
    \includegraphics[width=0.49\textwidth]{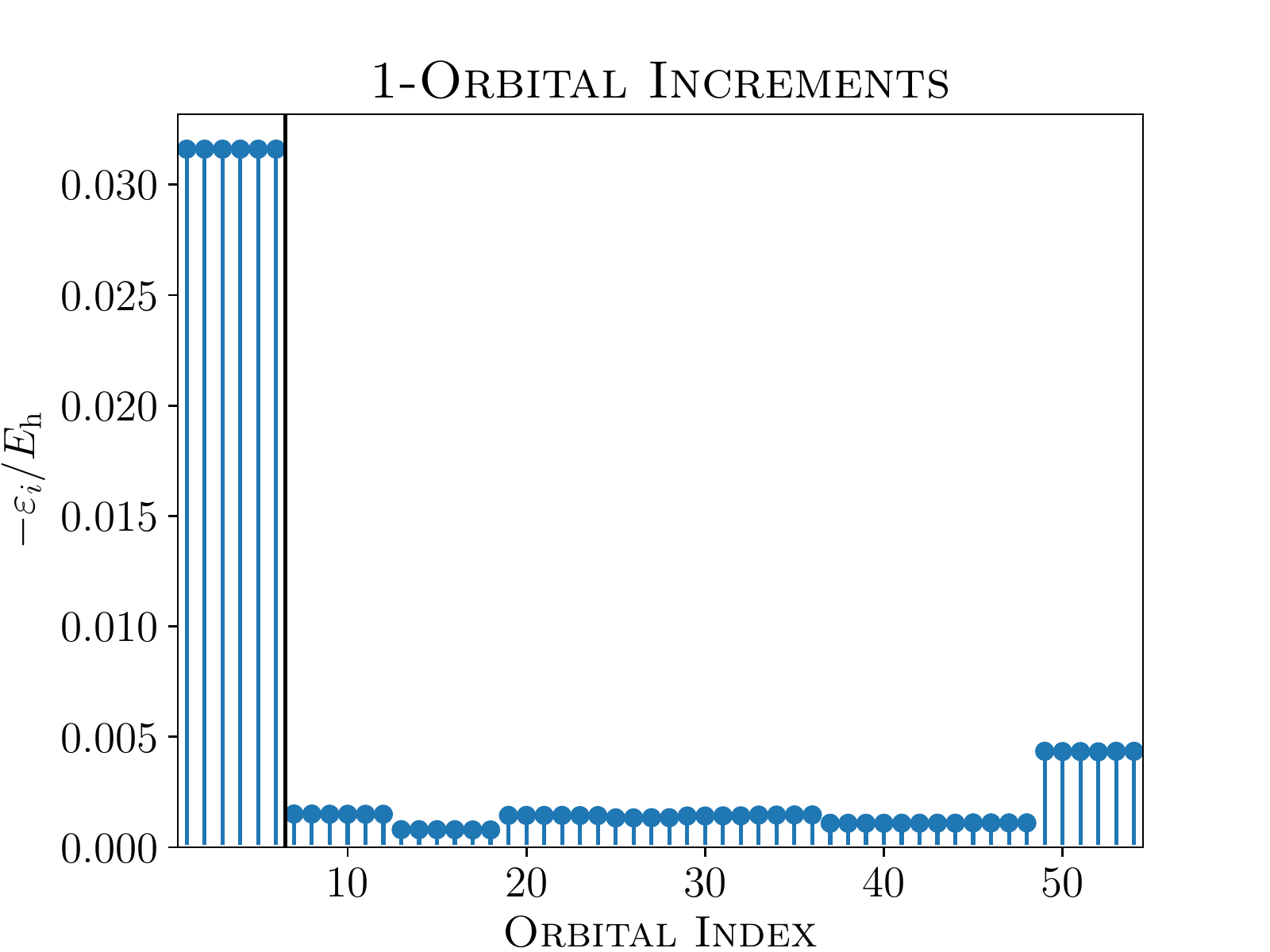}
    \includegraphics[width=0.49\textwidth]{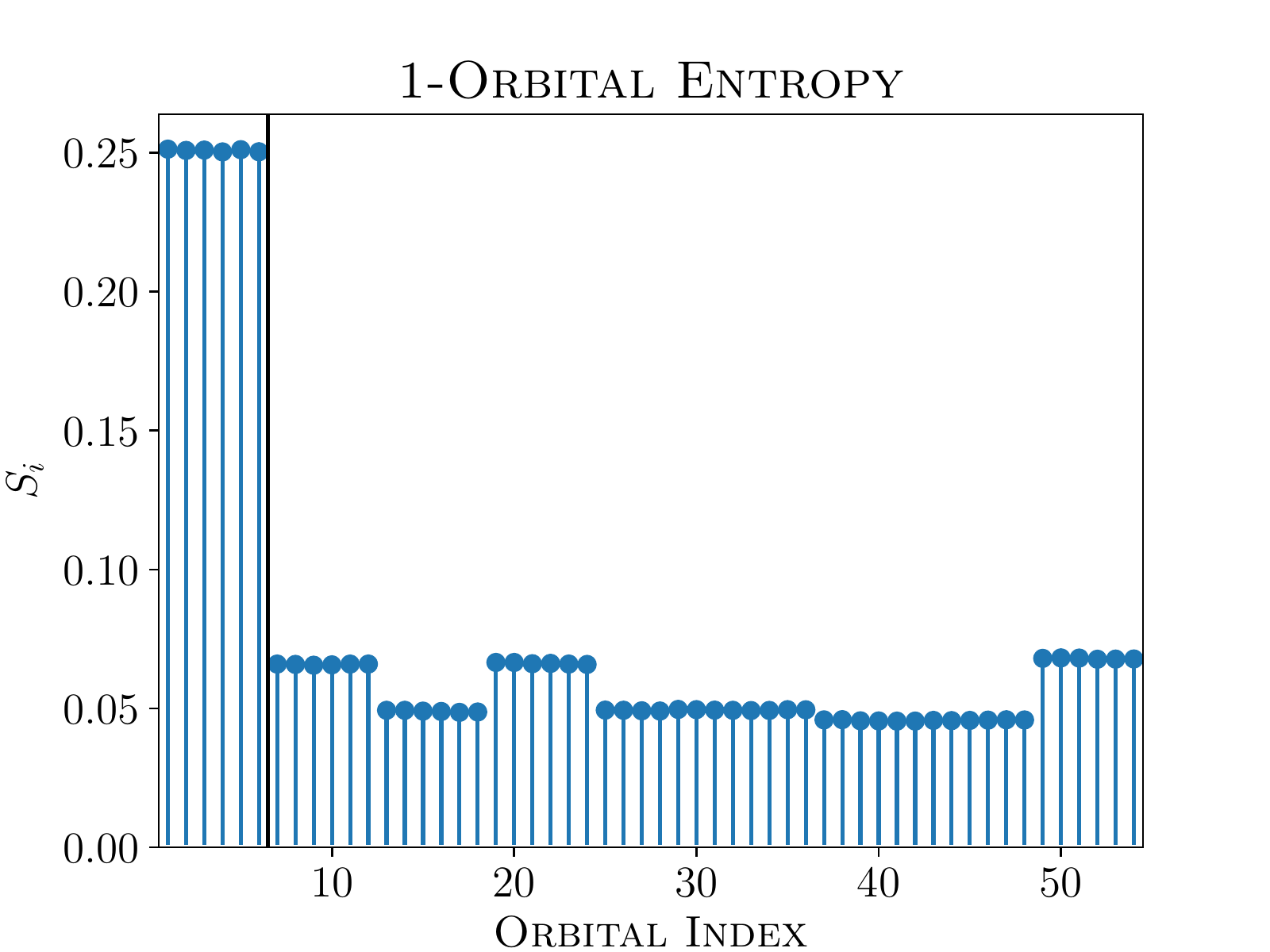}\\
    \includegraphics[width=0.49\textwidth]{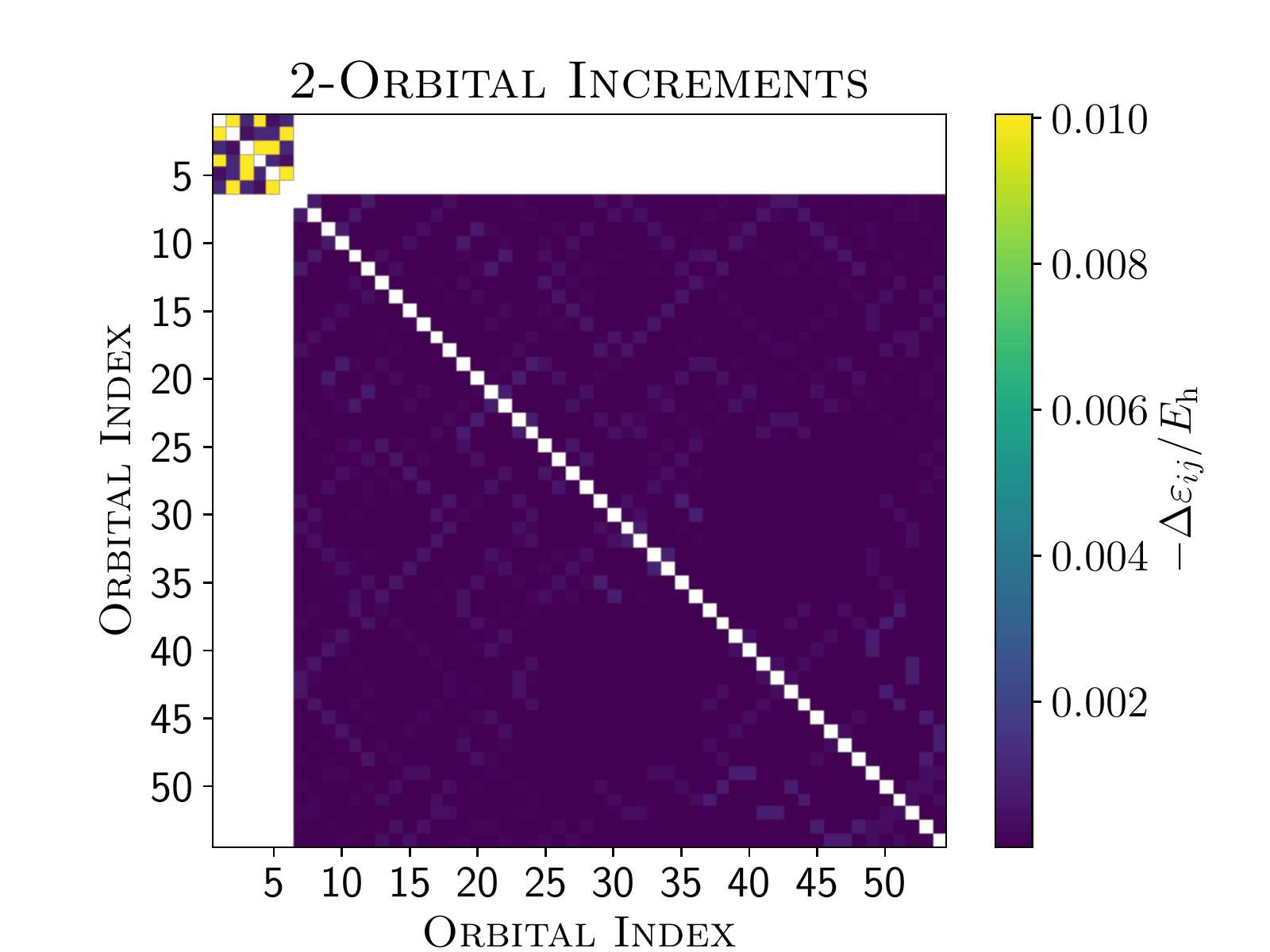}
    \includegraphics[width=0.49\textwidth]{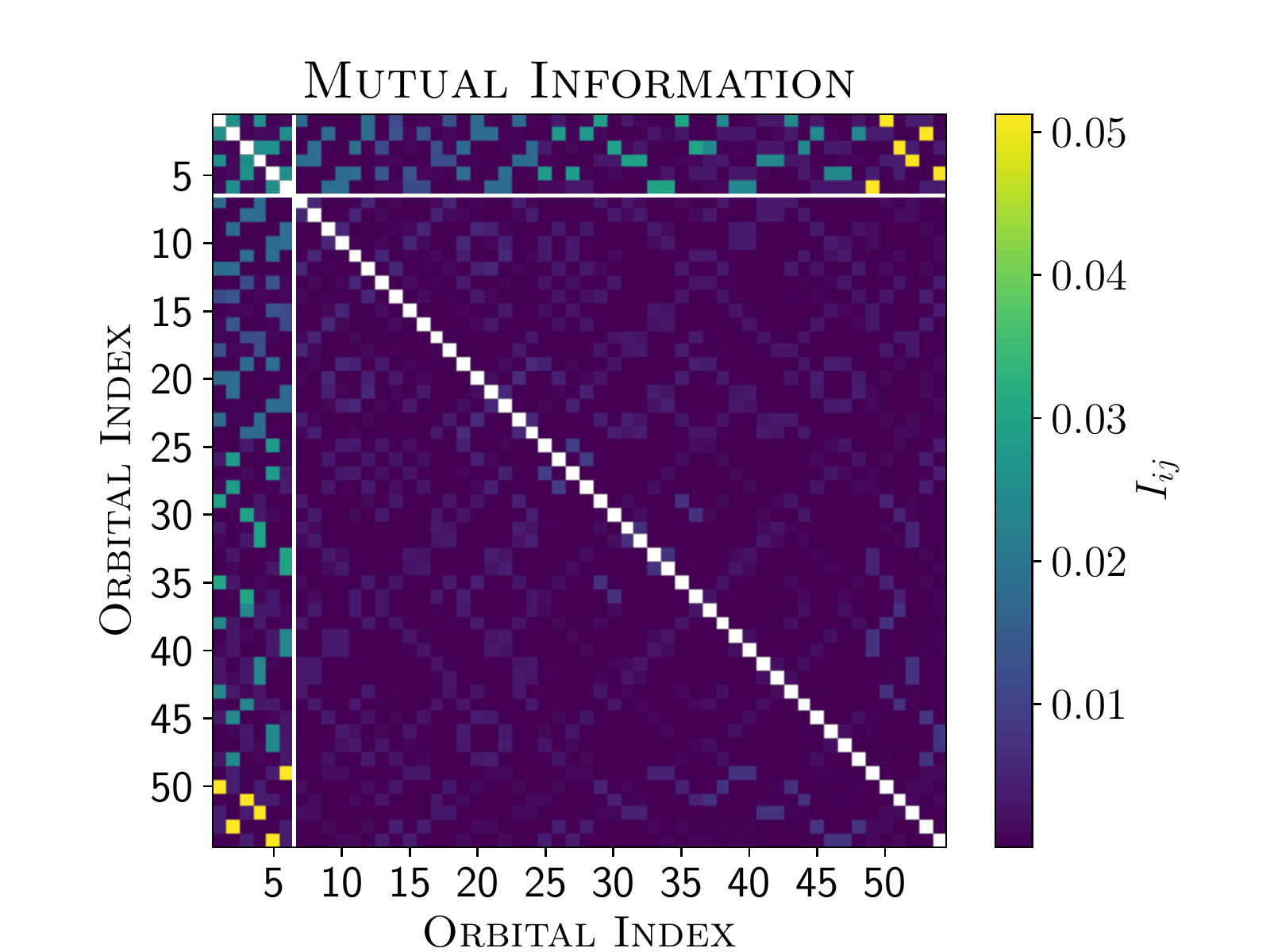}\\
    \includegraphics[width=0.49\textwidth]{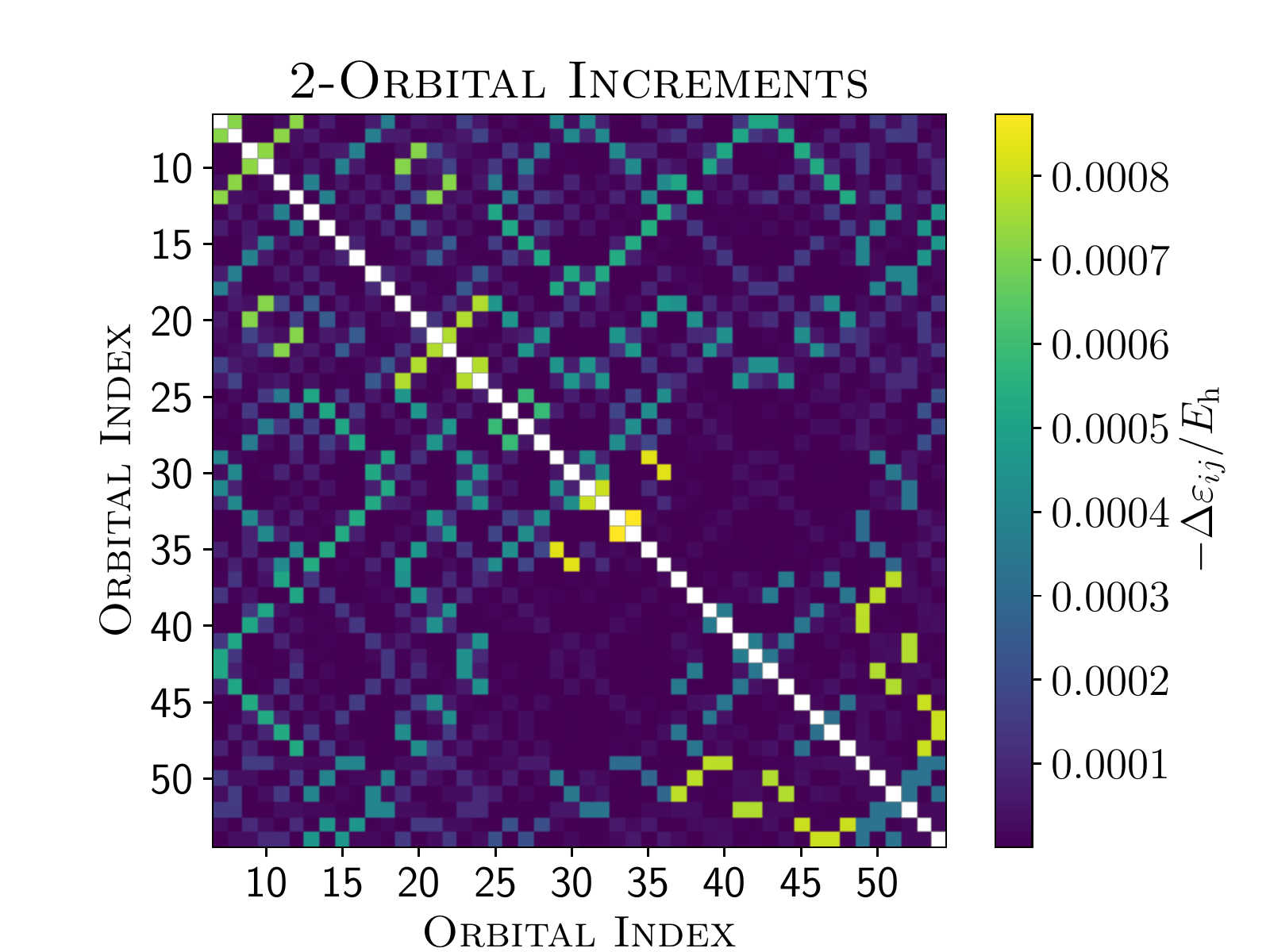}
    \includegraphics[width=0.49\textwidth]{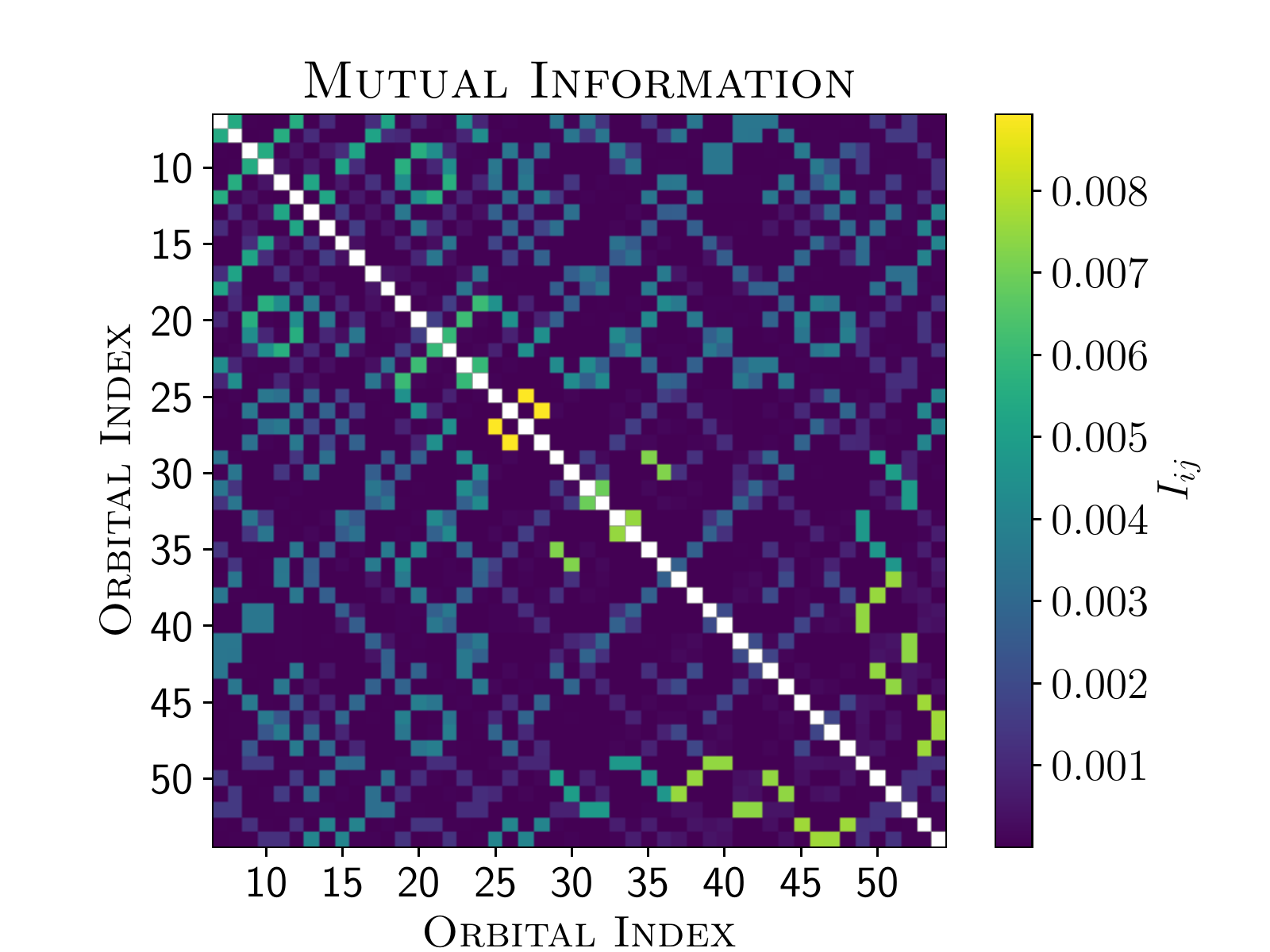}\\
    \caption{\ce{Be6} at equilibrium distance ($R=\SI{2.2}{\angstrom}$). Comparison of Increments and Orbital Entropies for quantifying orbitals correlations. Left and right column show results based on Methods of Increments (MoI) and Quantum Information Theory (QIT) respectively. The upper and middle row show 1-orbital and 2-orbital correlations respectively. The Highest Occupied Molecular Orbital (HOMO) and Lowest Unoccupied Molecular Orbital (LUMO) are separated by black and white lines respectively. In the lower row the 2-orbital correlations among virtual orbitals are shown. All orbitals are ordered by increasing diagonal Fock matrix element.}
    \label{fig:Be6_conf1}
\end{figure}

\begin{figure}
    \centering
    \includegraphics[width=0.49\textwidth]{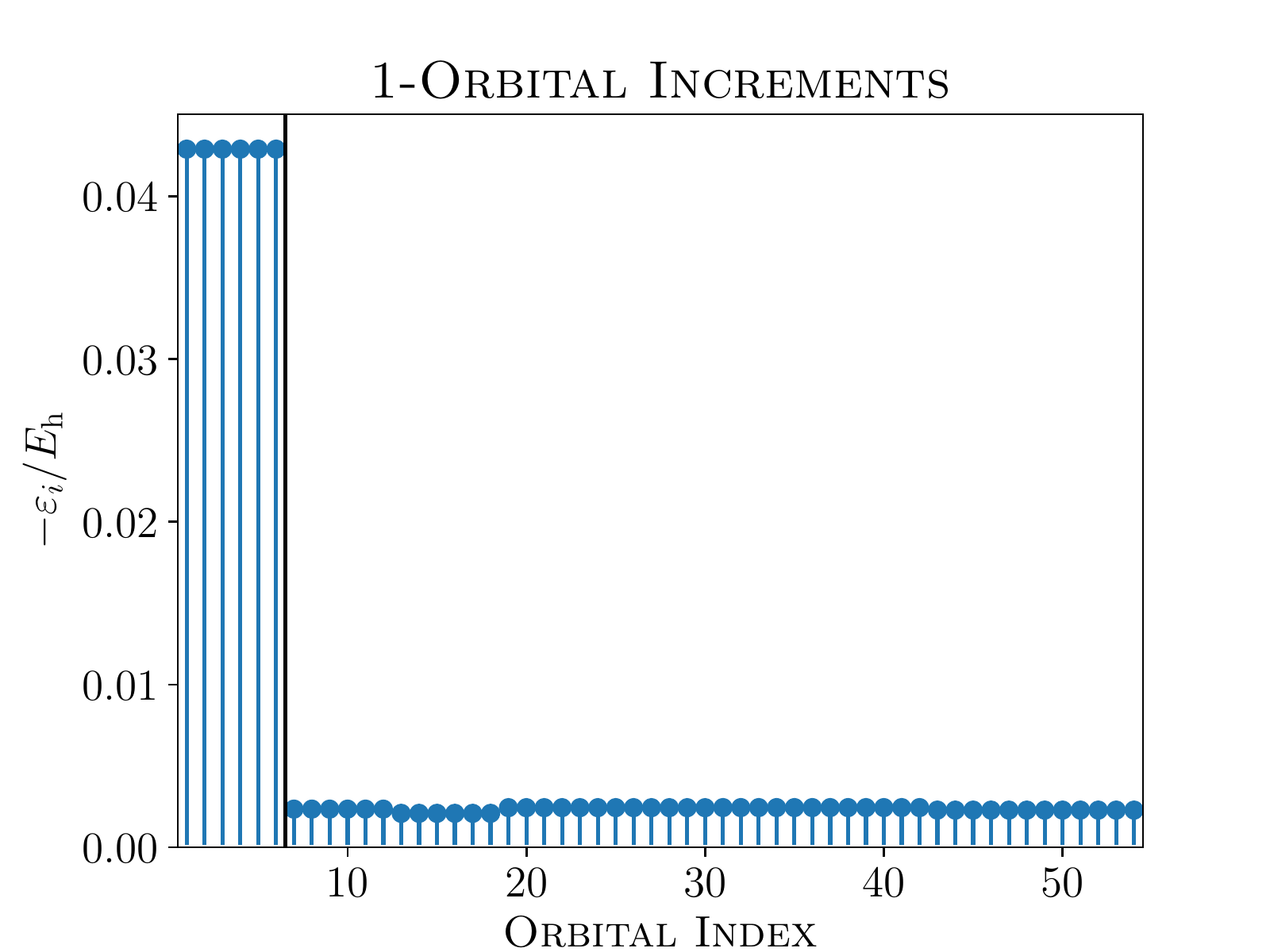}
    \includegraphics[width=0.49\textwidth]{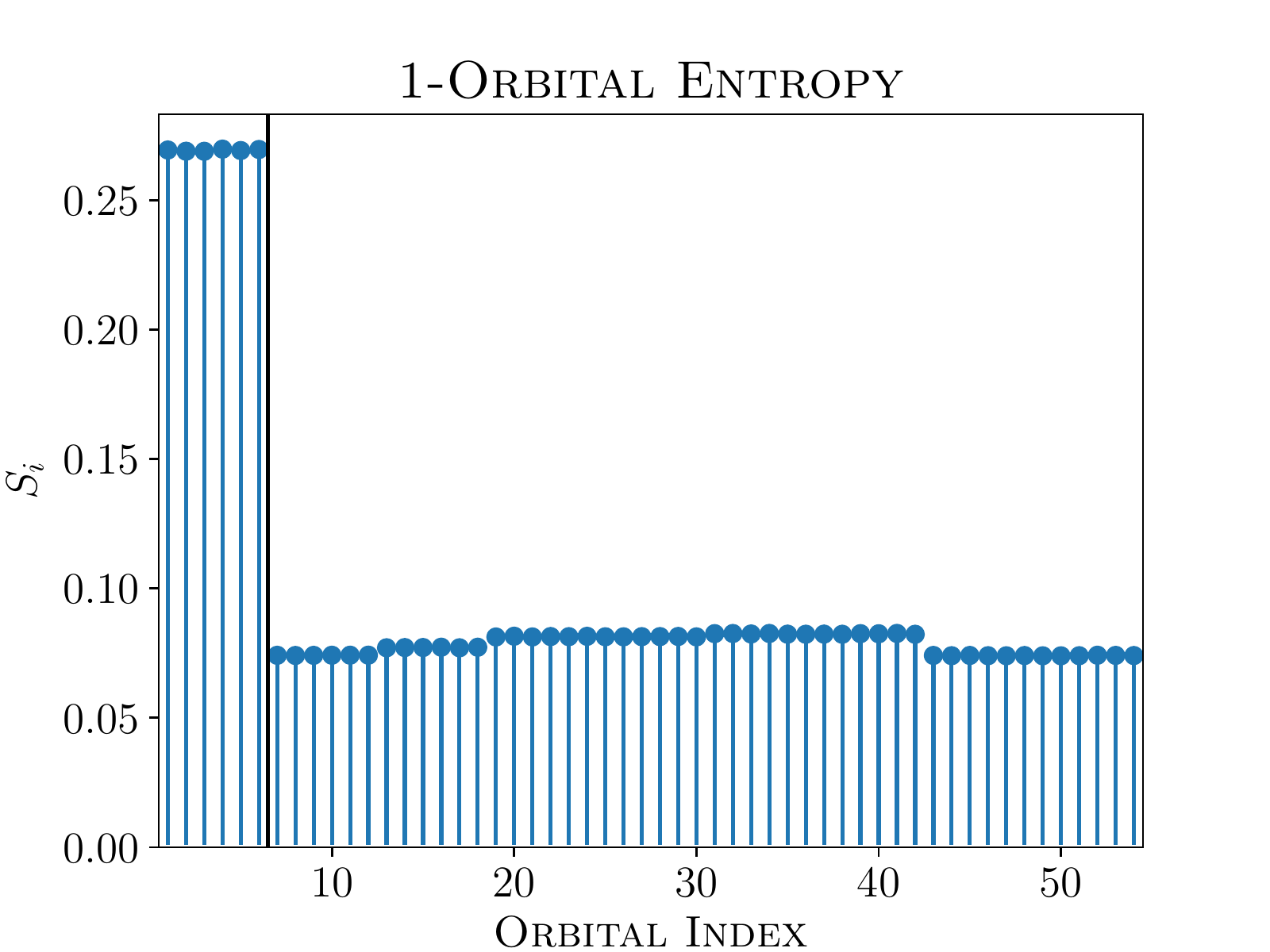}\\
    \includegraphics[width=0.49\textwidth]{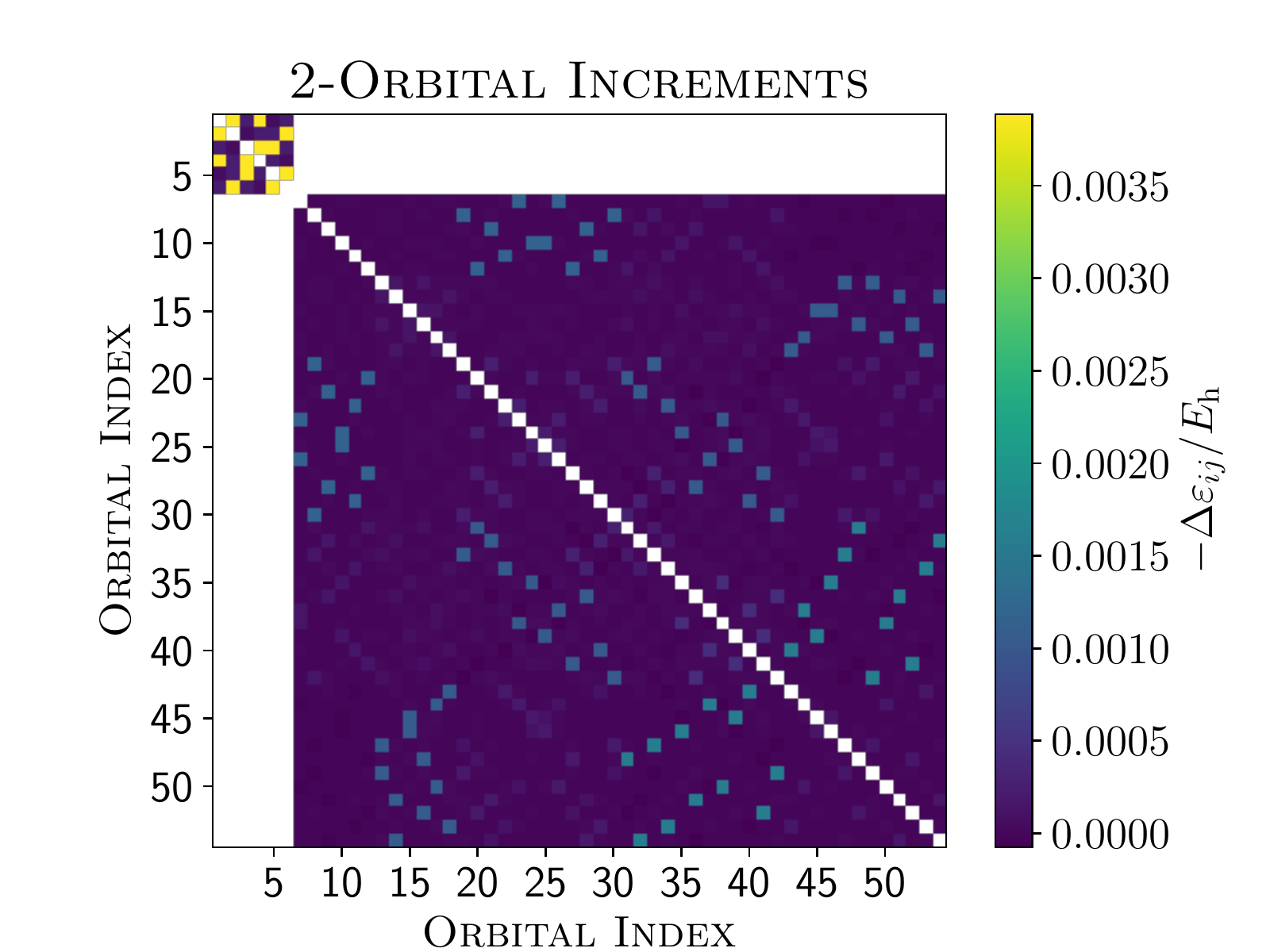}
    \includegraphics[width=0.49\textwidth]{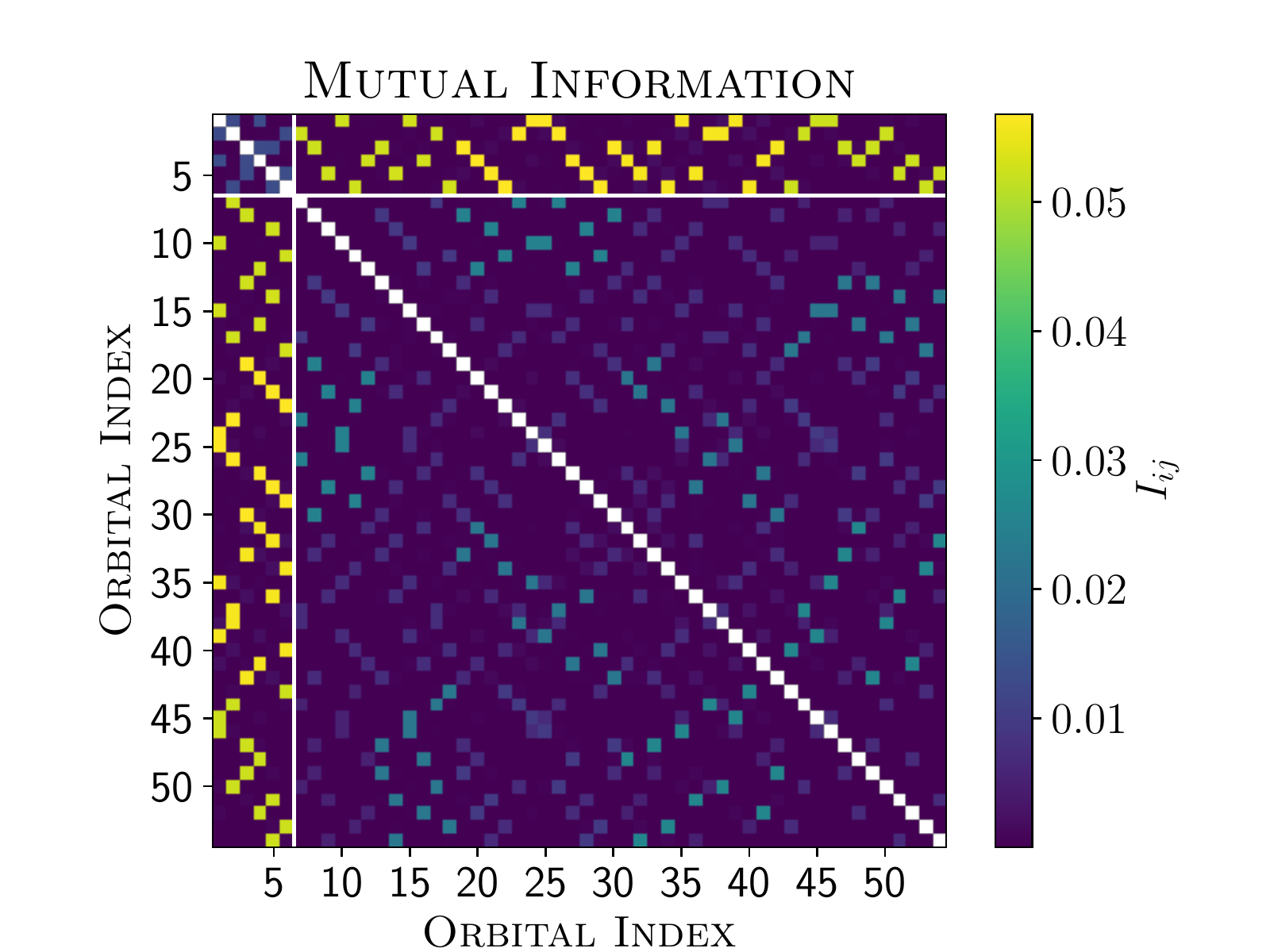}\\
    \includegraphics[width=0.49\textwidth]{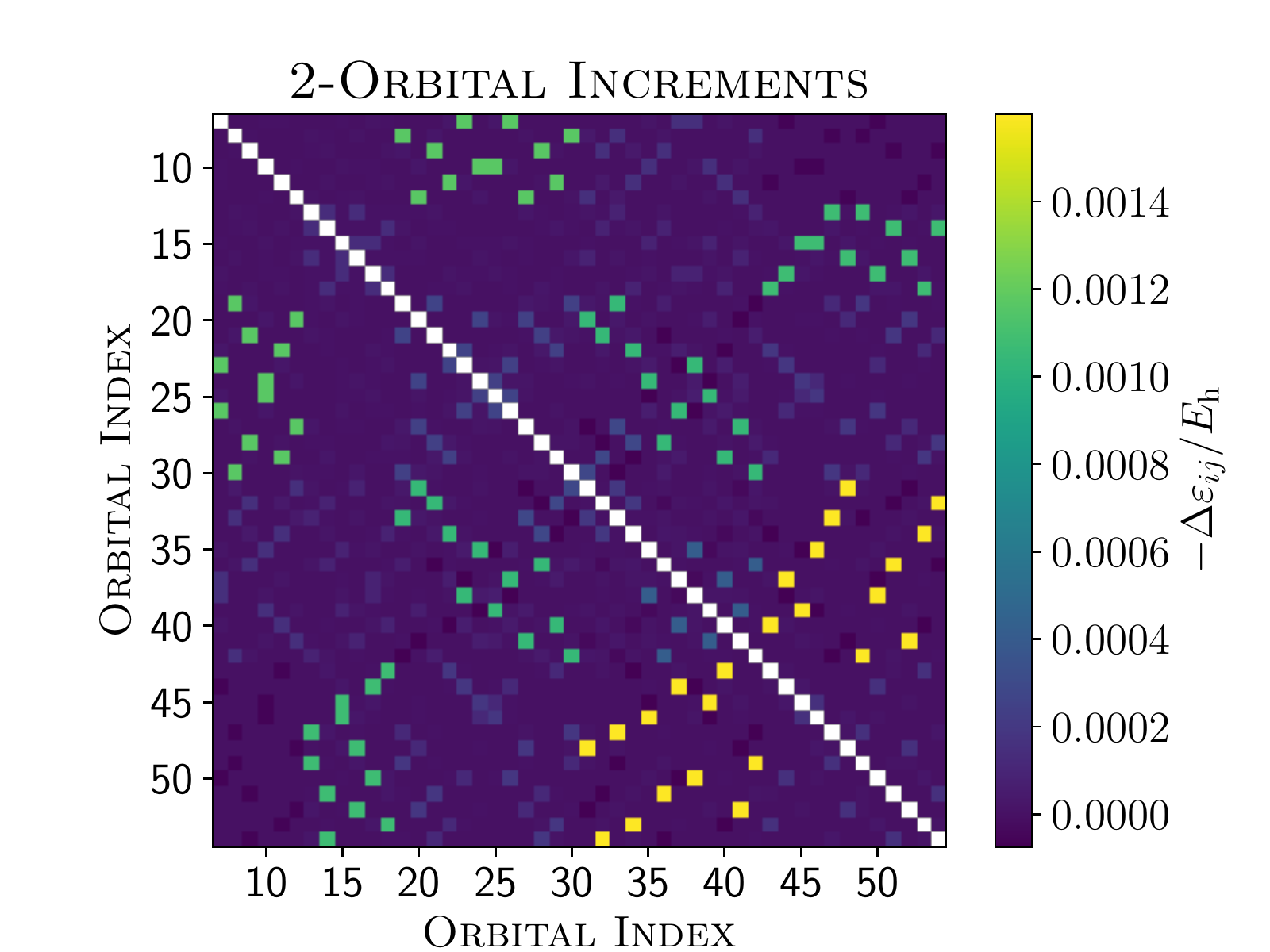}
    \includegraphics[width=0.49\textwidth]{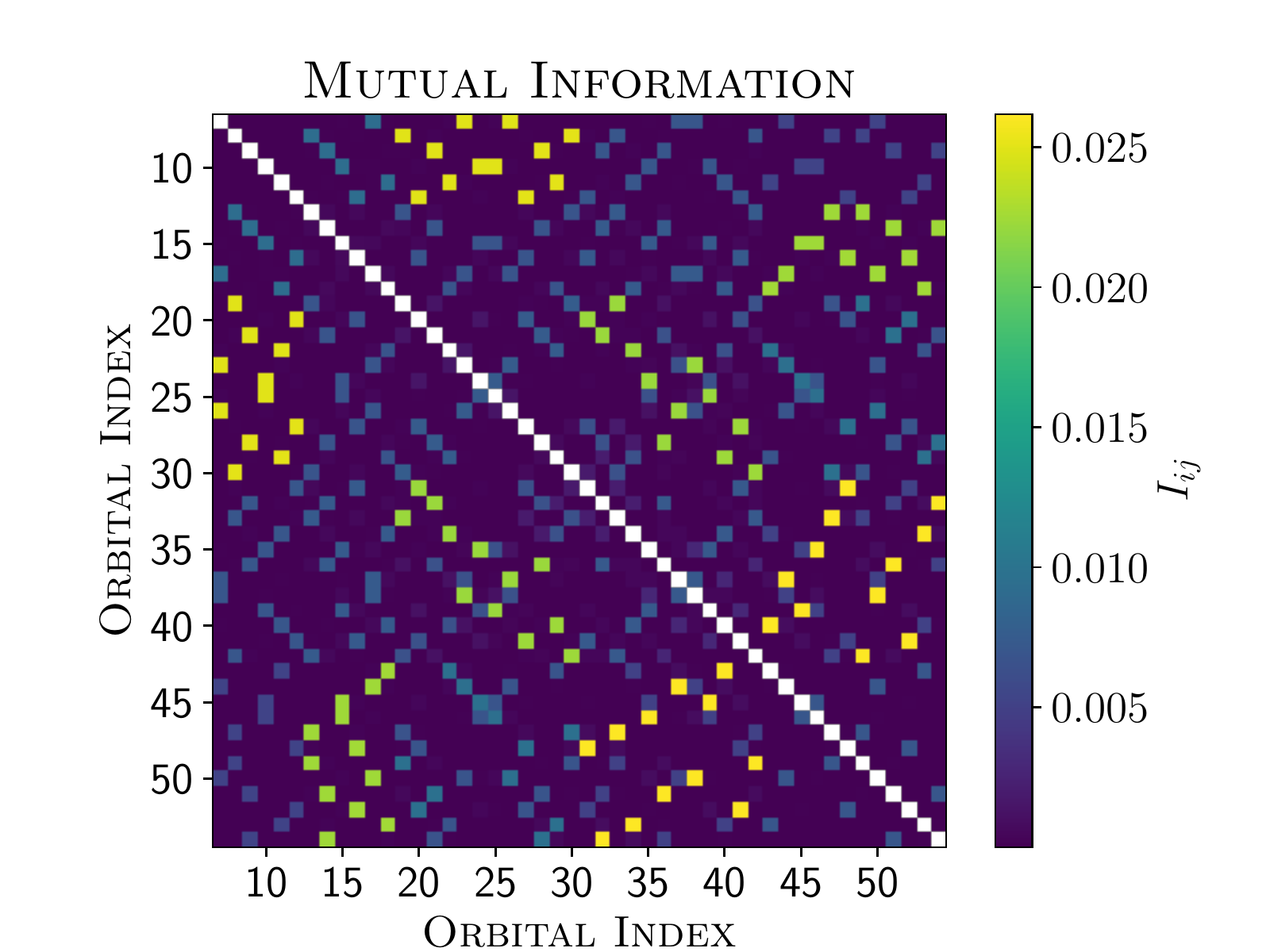}\\
    \caption{\ce{Be6} at dissociation limit ($R=\SI{3.5}{\angstrom}$). Comparison of Increments and Orbital Entropies for quantifying orbitals correlations. Left and right column show results based on Methods of Increments (MoI) and Quantum Information Theory (QIT) respectively. The upper and middle row show 1-orbital and 2-orbital correlations respectively. The Highest Occupied Molecular Orbital (HOMO) and Lowest Unoccupied Molecular Orbital (LUMO) are separated by black and white lines respectively. In the lower row the 2-orbital correlations among virtual orbitals are shown. All orbitals are ordered by increasing diagonal Fock matrix element.}
    \label{fig:Be6_conf2}
\end{figure}

The 1-orbital increments for all virtual orbitals are shown in \cref{fig:Be6_allvirtual}. For both cases, equilibrium distance and dissociation limit, we can easily identify a set of 24 orbitals with values close to zero. This allows for a clear cut and we select the the remaining 48 virtual orbitals for the active space. Together with the six occupied orbitals this results in a (12,54) active space.
All MoI and DMRG results are presented in \cref{fig:Be6_conf1} for the equilibrium distance and \cref{fig:Be6_conf2} for the dissociated situation. The increments have been calculated for the full virtual orbital space (12,78). However, the virtual orbital increments are identical for both active spaces, and the occupied increments will only change a little, as we only removed those virtual orbitals which have negligible correlation effects.\\

Calculated energies for the different active spaces are summarized in \cref{tab:Be6} and agree with previously reported results \cite{Koch-2016}.
The increments and QIT data are again in a rough agreement. Obvious difference, for example the 1-orbital increment and 1-orbital entropy at equilibrium distance (cf. \cref{fig:Be6_conf1}), are in the order of magnitude to be expected based on the previous discussion for the polyacetelene system. Comparison with QIT results for neglecting virtual orbitals based on very small 1-orbital increments was not possible, as DMRG is not feasible to treat dynamical correlation. However, the 2-orbital data for the remaining virtual orbitals is in good agreement.\\

\begin{table}
    \centering
    \caption{Total energy and correlation energy for a \ce{Be6} ring obtained with different methods using a cc-pVDZ basis set. All energies are in \si{\hartree}.}
    \label{tab:Be6}
    \begin{ruledtabular}
    \begin{tabular}{cccccc}
        & \multicolumn{2}{c}{Total Energy} & \multicolumn{2}{c}{Correlation Energy} & Dissociation Energy \\
        & $R=\SI{2.2}{\angstrom}$ & $R=\SI{3.5}{\angstrom}$ & $R=\SI{2.2}{\angstrom}$ & $R=\SI{3.5}{\angstrom}$ \\
        \toprule
        HF          & $-87.573755$ & $-87.421411$ & & & $0.152344$ \\
        CCSD(T) (canonical) & $-87.828746$ & $-87.701393$ & $-0.254991$ & $-0.279981$ & $0.127353$ \\
        DMRG(12,54) & $-87.777764$ & $-87.644043$ & $-0.204009$ & $-0.222632$ & $0.133721$ \\
        CAS(12,78)-MoI (occupied) & $-87.831646$& $-87.703778$ & $-0.257890$ & $-0.282367$ & $0.127868$ \\
         1-orbital & & & $-0.189642$ & $-0.257473$ \\
         2-orbital & & & $-0.068249$ & $-0.024894$ \\
        CAS(12,78)-MoI (virtual) & $-87.824877$ & $-87.694914$ & $-0.251122$ & $-0.273503$ & $0.129963$ \\
         1-orbital & & & $-0.082232$ & $-0.123687$ \\
         2-orbital & & & $-0.131174$ & $-0.091626$ \\
         3-orbital & & & $-0.037715$ & $-0.058190$ \\
    \end{tabular}
    \end{ruledtabular}
\end{table}

As a reference for the total energy we use CCSD(T) based on the canonical orbitals. Being a single-reference approach, it cannot fully recover strong correlation effects. However, we do get agreement up to a couple of \si{\milli\hartree} with the CAS(12,78)-MoI. Expanding in terms of occupied orbitals does yield good convergence already at the 2-orbital increment level, while the virtual orbital expansion requires 3-orbital increments. The latter is much easier to calculate due to the much smaller active space of each individual increment, at the cost of having a much higher number of increments.
The gain of a smaller active space outweighs the increasing number of individual increments.
As the calculation of the 3-orbital increments for the occupied orbital expansion is neither feasible nor necessary we omit them here.
Lastly, the DMRG calculations yield larger total energies and has slightly larger dissociation energy, due to the missing dynamical correlation.

\section{Summary and Discussion}
\label{sec:summary}

We have performed calculation recovering static and dynamical correlations by applying the MoI and DMRG. By comparing the individual contribution (increments vs. entropies) we can see, that both show very similar patterns in terms of which contributions are most important. The actual values however, show large difference, as was expected due to the different nature of increments and entropies.
This has been shown for two different model systems, polyacetelene and \ce{Be6} rings.
Deducing from one set of results to the other thus seems an appropriate approach, care must be taken though when deciding on which cutoff parameter to use when neglecting contributions.\\

Comparing the computational effort for both methods we see that MoI is rather restricted. Each individual increment requires a separate calculation, and the number of increments increases combinatorially with the number of orbitals (centers) and level. Furthermore, MoI it is not capable of showing cross-correlations between occupied and virtual orbitals. This can be problematic when trying to determine a smaller appropriate active space.
Here QIT has the advantage of providing all relevant information, just based on a single many-body wave function. Additionally, QIT can be easily extended to consider contributions from combining 3, 4 or more groups since the wave function is readily available.
As a further benefit the QIT entropies are insensitive to the quality of the DMRG calculation (number of block states). Indeed the one-orbital entropies and mutual information presented here (cf. \cref{fig:MoIvsQIT}) can be obtained with same quality by using only $M=128$ block states (cf. Fig.~S5 of the Supplementary Information). The time and memory limiting factor is then likely to be the construction and diagonalization of the reduced density matrices.\\

As possible applications we may first calculate all 1-orbitals increments to construct a suitable active space for DMRG, as we already applied here for the \ce{Be6} ring. This exploits the cheap 1-orbital increment calculations.
On the other hand, screening all static correlation effects by applying DMRG can be used to select relevant higher order increments, which are rapidly increasing in computational effort.\\


\begin{acknowledgments}
    We would like to thank Prof. Peter Fulde (Dresden) for discussion and stimulating this topic. Financial support from the International Max Planck Research School ``Functional Interfaces in Physics and Chemistry'' is gratefully acknowledged. The high performance computing facilities of the Freie Universität Berlin (ZEDAT) are acknowledged for computing time.
\end{acknowledgments}

\bibliography{DMRG_QIT.bib,MoI.bib,molpro.bib,methods.bib}

\end{document}